\PassOptionsToPackage{pdfpagelabels=false}{hyperref}
\documentclass[fleqn,usenatbib]{mnras}

\usepackage[T1]{fontenc}
\usepackage{ae,aecompl}

\usepackage{float}
\usepackage{hyperref}
\usepackage{apjfonts}
\usepackage{amssymb}
\usepackage{amsmath}
\usepackage{graphicx}	
\usepackage[normalem]{ulem}
\usepackage{makecell}
\usepackage{mathtools}
\usepackage{dutchcal}
\usepackage{supertabular}
\usepackage[all]{hypcap}
\usepackage{array,booktabs,multirow}
\usepackage{tabulary}
\hypersetup{colorlinks,citecolor=blue}
\usepackage{etoolbox}
\makeatletter
\patchcmd\@combinedblfloats{\box\@outputbox}{\unvbox\@outputbox}{}{%
  \errmessage{\noexpand\@combinedblfloats could not be patched}%
}%
\makeatother

\newcommand{\mystar}{\ast}        
\newcommand\run[1]{\texttt{#1}}   

\newcommand{\vmax}{V_\mathrm{max}}

\newcommand{\msun}{\,\textnormal{M}_\odot}

\newcommand{\mstar}{M_\mystar}

\newcommand{\mhalo}{M_\mathrm{halo}}

\newcommand{\mpc}{\mathrm{Mpc}}

\newcommand{\kpc}{\mathrm{kpc}}

\newcommand{\lcdm}{$\Lambda$CDM}

\newcommand{\etal}{et al.}

\newcommand{\gyr}{\mathrm{Gyr}}

\newcommand{\tten}{t_{10}}
\newcommand{\tfifty}{t_{50}}
\newcommand{\tnintey}{t_{90}}

\newcommand{\dhost}{d_{\rm host}}
\newcommand{\dnearest}{d_{\rm nearest}}
\newcommand{\mnearest}{M_{\rm halo}^{\rm nearest}}
\newcommand{\zreion}{z_\mathrm{reion}}

\newcolumntype{M}[1]{>{\centering\arraybackslash}m{#1}}

\newcommand\altaffilmark[1]{$^{#1}$}
\newcommand\altaffiltext[2]{$^{#1}$#2}

\title[Dwarf SFHs in FIRE-2]{
Star formation histories of dwarf galaxies in the FIRE simulations:  dependence on mass and Local Group environment
\vspace{-0.7cm}}

\vspace{-0.3cm}
\author[S. Garrison-Kimmel \etal]{
\parbox[t]{\textwidth}{
Shea Garrison-Kimmel\textsuperscript{\href{https://orcid.org/0000-0002-4655-8128}{\includegraphics[width=2.5mm]{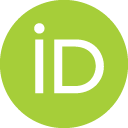}}\,}\thanks{$\!$sheagk@caltech.edu}\thanks{ Einstein Fellow}\altaffilmark{1},
Andrew Wetzel\textsuperscript{\href{https://orcid.org/0000-0003-0603-8942}{\includegraphics[width=2.5mm]{orcid.png}}\,}\altaffilmark{2},
Philip F. Hopkins\altaffilmark{1},
Robyn Sanderson\altaffilmark{1,3,4},
Kareem El-Badry\altaffilmark{5},
Andrew Graus\altaffilmark{6,7},
T.K. Chan\altaffilmark{8},
Robert Feldmann\textsuperscript{\href{https://orcid.org/0000-0002-1109-1919}{\includegraphics[width=2.5mm]{orcid.png}}\,}\altaffilmark{9},
Michael Boylan-Kolchin\altaffilmark{7},
Christopher Hayward\altaffilmark{4},
James S. Bullock\altaffilmark{6},
Alex Fitts\altaffilmark{7},
Jenna Samuel\altaffilmark{2},
Coral Wheeler\altaffilmark{1},
Du{\v s}an Kere{\v s}\altaffilmark{8}, and
Claude-Andr{\'e} Faucher-Gigu{\`e}re\altaffilmark{10}
}
\vspace*{10pt} \\
\parbox{\textwidth}{
\footnotesize
\altaffiltext{1}{TAPIR, Mailcode 350-17, California Institute of Technology, Pasadena, CA 91125, USA} \\
\altaffiltext{2}{Department of Physics, University of California, Davis, CA 95616, USA} \\
\altaffiltext{3}{Department of Physics \& Astronomy, University of Pennsylvania, 209 South 33rd St., Philadelphia, PA 19104 USA} \\
\altaffiltext{4}{Center for Computational Astrophysics, Flatiron Institute, 162 5th Ave., New York, NY 10010 USA} \\
\altaffiltext{5}{Department of Astronomy and Theoretical Astrophysics Center, University of California Berkeley, Berkeley, CA 94720} \\
\altaffiltext{6}{Center for Cosmology, Department of Physics and Astronomy, University of California, Irvine, CA 92697, USA} \\
\altaffiltext{7}{Department of Astronomy, The University of Texas at Austin, 2515 Speedway, Stop C1400, Austin, TX 78712, USA} \\
\altaffiltext{8}{Department of Physics, Center for Astrophysics and Space Science, University of California at San Diego, 9500 Gilman Drive, La Jolla, CA 92093} \\
\altaffiltext{9}{Institute for Computational Science, University of Zurich, CH-8057 Zurich, Switzerland} \\
\altaffiltext{10}{Department of Physics and Astronomy and CIERA, Northwestern University, 2145 Sheridan Road, Evanston, IL 60208, USA} \\
\vspace{-0.75cm}}
}

\date{Accepted XXX. Received YYY; in original form ZZZ}

\pubyear{2019}

\begin{document}
\label{firstpage}
\pagerange{\pageref{firstpage}--\pageref{lastpage}}

\maketitle 
\begin{abstract}
We study star formation histories (SFHs) of $\simeq500$ dwarf galaxies 
(stellar mass $\mstar=10^5 - 10^9\msun$) from FIRE-2 cosmological zoom-in 
simulations.  We compare dwarfs around individual Milky Way (MW)-mass 
galaxies, dwarfs in Local Group (LG)-like environments, and true field 
(i.e. isolated) dwarf galaxies.  We reproduce observed trends wherein 
higher-mass dwarfs quench later (if at all), regardless of environment.
We also identify differences between the environments, both in terms of 
``satellite \emph{vs.} central'' and ``LG \emph{vs.} individual MW 
\emph{vs.} isolated dwarf central.''
Around the individual MW-mass hosts, we recover the result expected from 
environmental quenching: central galaxies in the ``near field'' 
have more extended SFHs than their satellite counterparts, with the former 
more closely resemble isolated (``true field'') dwarfs (though near-field 
centrals are still somewhat earlier forming).  However, this difference is muted in the LG-like 
environments, where both near-field centrals and satellites have similar SFHs, which 
resemble satellites of single MW-mass hosts.  This distinction is 
strongest for $\mstar=10^6$--$10^7\msun$ but exists at other masses.    
Our results suggest that the paired halo nature of the LG may regulate star 
formation in dwarf galaxies even beyond the virial radii of the MW and 
Andromeda. Caution is needed when comparing zoom-in 
simulations targeting isolated dwarf galaxies against observed dwarf 
galaxies in the LG.
\end{abstract}




\section{Introduction}
\label{sec:intro}
The star formation history (SFH) of a dwarf galaxy is one of its 
fundamental properties.  It has implications for the $z=0$ dark 
matter density profile at fixed mass, as late time star formation
appears to correlate with core formation \citep[e.g.][]{Onorbe2015,
Read2016,Read2018}.  Dwarf SFHs further inform how interactions with 
the Milky Way (MW) impact satellites, either through comparisons with 
infall times inferred for individual satellites (e.g. \citealp{Rocha2012}; Fillingham et al., in preparation)
or more broadly in comparing typical satellite SFHs with those of 
central (non-satellite) galaxies, as suggested by \citet{BrooksZolotov2012}.  
They can also yield unique constraints on the contribution of dwarf 
galaxies to the reionizing background \citep[e.g.][]{Ricotti2008,Weisz2014b,
Boylan-Kolchin2015}.  Moreover, the shape of a dwarf's SFH may be 
correlated with the growth of the halo at low halo masses 
($\simeq10^{10}\msun$; e.g. \citealp{Fitts2017}) and with the 
kinematics of the gas and stars in the galaxy at slightly higher 
masses ($\simeq10^{11}\msun$; e.g. \citealp{ElBadry2018}). Dwarf 
SFHs may even inform the nature of dark matter, as different DM 
models predict different accretion histories
\citep[e.g.][]{Governato2015,Colin2015,Lovell2017,Bozek2018}.

Observations have begun to provide detailed constraints on the star 
formation histories of a large fraction of the dwarf galaxies in the 
Local Group (LG; defined as the cosmological volume containing the MW, 
M31, and all dwarf galaxies within $\sim1~\mpc$ of either of these 
hosts), typically by resolving the oldest main-sequence turn off stars with 
space-based photometry.  \citet{WeiszSFH} presented the largest such 
sample, with SFHs for forty LG dwarf galaxies uniformly derived from 
HST observations \citep[and also see][]{Cole2007,Cole2014,Skillman2017}.  
They found that higher mass galaxies form a higher fraction of their 
stars at later times, and that the central galaxies in the so-called 
Local Field (i.e., more than $300~\kpc$ from both the MW or M31, but 
still within the Local Group) typically form later than their 
satellite counterparts.  \citet{Weisz2014:ReionSigs} argued that only 
two of those forty dwarf galaxies are consistent with their star formation 
being completely halted (quenched) by reionization, though \citet{Brown2014} 
used similar observations of lower mass ultra-faint dwarf galaxies (stellar 
mass $\mstar\lesssim3\times10^4\msun$) to argue that reionization becomes 
increasingly important at lower masses, as all six of the galaxies in their 
sample stop forming stars by $z\simeq2$.  \citet{RodriguezWimberly2018}
further argued that environmental effects caused by the hot halo of the 
MW \citep[e.g.][]{Gupta2012}~--~including ram-pressure and turbulent 
viscous stripping \citep{Gunn1972,Nulsen1982,Hester2006,Fillingham2016}, 
which actively remove gas from satellite galaxies, and ``starvation,'' 
where the accretion of fresh gas is suppressed \citep{Larson1980,
Kawata2008}~--~cannot reproduce such early quenching times, strengthening 
the case that the UV background is responsible.

At higher masses, however, it appears that environment is more important
in shaping dwarf SFHs, as epitomized by the simple observation that 
the majority of satellite dwarfs have no detectable HI while the majority
of centrals do \citep[e.g.][]{Einasto1974,McConnachie2012,Spekkens2014}.  
The concept of environmental quenching is supported by the results of 
\citet{Geha2012}, who showed that the fraction of quenched 
$\mstar=10^8$--$10^9\msun$ galaxies is consistent with zero
for dwarfs $\gtrsim1~\mpc$ from the nearest MW-mass (or larger) 
host, but rises sharply at smaller distances.  Similarly, 
\citet{Gallart2015} argued that dwarf galaxies in the LG can 
be grouped into ``fast'' and ``slow'' rising SFHs, where the 
former have positions and/or velocities consistent with previous 
interactions with the MW or M31.  Several authors have used statistical arguments to show 
that the preponderance of quenched satellites around the MW 
suggests that quenching times must be quite short at low masses 
($\lesssim2~\gyr$ for $\mstar \lesssim 10^8\msun$), and longer 
($\gtrsim6~\gyr$) at higher masses, provided that quenching 
is directly linked with entering the virialized volume of the MW
\citep[e.g.][]{Wetzel2013,Wheeler2014,Fillingham2015,Wetzel2015b}.
\citet{Fillingham2018} recently used a similar technique to argue
that quenching processes that operate purely within the virial radius 
of the hosts are insufficient to explain the quenched galaxies at 
$\gtrsim600~\kpc$ from the MW/M31, though dwarf-dwarf interactions
may lead to more extended (and perhaps more slowly collapsing) HI 
reservoirs relative to isolated dwarf galaxies \citep{Pearson2016}.
\citet{Emerick2016} used idealized wind-tunnel simulations to argue
that stripping, aided by supernovae feedback, could not explain the
short quenching timescales inferred by the above works even for 
satellites.

The question of satellite quenching is further complicated 
by the results of the Satellites Around Galactic Analogs (SAGA) Survey:  
26 of the 27 satellites identified around the eight MW-mass galaxies 
in their sample appear to be star forming, though the 
dwarf sample only reaches $\mstar\gtrsim10^7\msun$ \citep{SAGA1}.  In slight contrast, 
\citet{Tanaka2018} reached $\mstar\gtrsim10^6\msun$ and identified a 
mix of blue and red satellite galaxies around two other MW analogs, 
with the authors classifying the majority of the blue (i.e. star-forming) sample as 
``possible'' dwarf galaxies, rather than secure detections.

Meanwhile, as the typical resolutions of hydrodynamic cosmological 
simulations increase, authors have begun to compare simulated dwarf
SFHs to the observations detailed above.  Many of these works have 
focused on highly isolated (i.e. field) dwarf galaxies, which can be 
simulated at much higher resolutions than dwarfs around MW-mass hosts 
because the MW-mass galaxy itself dominates the run-time of the latter 
such simulations.  For example, \citet{Fitts2017} presented dwarf SFHs 
taken from simulations using the FIRE\footnote{\url{http://fire.northwestern.edu}} 
(Feedback In Realistic Environments) physics.  They demonstrated overall
agreement with observations in terms of the range of dwarf SFHs.
They further argued that the $z=0$ stellar mass scales with the 
maximum circular velocity $\vmax$ of the halo (at fixed halo mass), 
though they did not identify any clear trends with the shapes of 
the SFHs.  \citet{Wright2018} used GASOLINE \citep{Wadsley2004} 
simulations of similarly isolated dwarf galaxies to understand why 
the SFHs of some dwarfs ``re-ignite'' after apparently quenching.  
They found that interactions with gaseous streams in the intergalactic 
medium can compress gas around the dwarf to the point where it begins 
to cool and form stars.  

Other authors have explored the SFHs of dwarf galaxies that evolve 
around MW-mass hosts, though typically at lower resolutions than the 
works above.  \citet{Benitez-Llambay2015} used the Constrained Local 
UniversE Simulations (CLUES; \citealp{Gottloeber2010}), which target LG-like pairs, to examine 
the impact of reionization on central galaxies.  They found great 
diversity in their simulated SFHs, and argued that ``gaps'' in 
star formation at intermediate ages (cosmic time $t\simeq4$--$8~\gyr$) 
can be attributed to reionization.  Similar to \citet{Fitts2017}, they 
argued for the importance of $\vmax$ in halos near the reionization 
suppression scale.  More recently, \citet{Wetzel2016} showed that 
the FIRE-2 prescriptions accurately reproduce the diversity in 
observed dwarf SFHs, the high fraction of quenched satellites 
near the MW at $z=0$, and the general dependence on galaxy mass,
though that paper examined only a single MW-mass host galaxy.  
Given the aforementioned results from SAGA and \citet{Tanaka2018}, 
it is unclear if the high quenched fraction around the MW is 
representative of MW-mass satellite populations, at least for 
$\mstar\gtrsim10^7\msun$.  Finally, \citet{Digby2018} examined the 
SFHs of dwarf galaxies in the APOSTLE \citep{Fattahi2016,Sawala2016} 
and AURIGA \citep{Grand2017} simulations.  They found that late-time 
($t\gtrsim8~\gyr$) star formation is suppressed in satellite galaxies 
relative to dwarf centrals of the same mass, and that low (high) mass 
dwarf centrals have SFHs that decline (rise) at late times.  
In a related work, \citet{Simpson2018} examined the gas content 
of dwarf galaxies in the AURIGA simulations.  Dwarf galaxies in 
their simulations are susceptible to ram pressure stripping, such 
that those at smaller host distances are more likely to be quenched
and gas poor.  At their lowest masses ($\mstar\simeq10^6\msun$), all 
dwarf galaxies that are either satellites today or were satellites in
the past are quenched, with the quenched fraction falling monotonically
with increasing stellar mass.

\begin{table}
\setlength{\tabcolsep}{3pt}
\begin{tabular}{lrrrrrr} 
Simulation     & $\mstar$             & $\mhalo$                    & $\dnearest$         & $\mnearest$                 & $m_i$                & Ref \\
               & $\left[\msun\right]$ & $\left[10^{10}\msun\right]$ & $\left[\mpc\right]$ & $\left[10^{10}\msun\right]$ & $\left[\msun\right]$ & \\ \midrule 
\multicolumn{7}{c}{\textit{Local Group hosts}} \\ 
\run{Romeo}    & $7.36\times 10^{10}$ & 132                         & 0.84                & 110.5                       & $3,500$              & A \\
\run{Juliet}   & $4.22\times 10^{10}$ & 110                         & 0.84                & 132.0                       & $3,500$              & A \\
\run{Thelma}   & $7.92\times 10^{10}$ & 143                         & 0.92                & 115.3                       & $4,000$              & A \\
\run{Louise}   & $2.85\times 10^{10}$ & 115                         & 0.92                & 143.3                       & $4,000$              & A \\
\multicolumn{7}{c}{\textit{Isolated MW hosts}} \\ 
\run{m12b}     & $9.42\times 10^{10}$ & 143                         & 3.99                & 37.8                        & $7,100$              & A \\
\run{m12c}     & $6.45\times 10^{10}$ & 135                         & 4.68                & 267.7                       & $7,100$              & A \\
\run{m12f}     & $8.78\times 10^{10}$ & 171                         & 3.91                & 76.2                        & $7,100$              & B \\
\run{m12i}     & $7\times 10^{10}$    & 118                         & 2.87                & 79.5                        & $7,100$              & C \\
\run{m12m}     & $1.26\times 10^{11}$ & 158                         & 3.94                & 279.3                       & $7,100$              & D \\
\run{m12r}     & $1.88\times 10^{10}$ & 110                         & 6.70                & 1064.6                      & $7,100$              & E \\
\run{m12w}     & $6.29\times 10^{10}$ & 108                         & 2.63                & 88.2                        & $7,100$              & E \\
\run{m12z}     & $2.25\times 10^{10}$ & 92.5                        & 3.37                & 34.4                        & $4,200$              & A \\ 
\multicolumn{7}{c}{\textit{Highly isolated dwarf centrals}} \\ 
\run{m10b}     & $4.68\times 10^{5}$  & 1.09                        & 4.78                & 38.9                        & $500$                & F \\
\run{m10c}     & $5.76\times 10^{5}$  & 1                           & 6.36                & 276.9                       & $500$                & F \\
\run{m10d}     & $1.56\times 10^{6}$  & 0.957                       & 9.09                & 115.8                       & $500$                & F \\
\run{m10e}     & $1.99\times 10^{6}$  & 1.17                        & 6.47                & 44.9                        & $500$                & F \\
\run{m10f}     & $4.2\times 10^{6}$   & 0.943                       & 4.83                & 55.0                        & $500$                & F \\
\run{m10g}     & $5.74\times 10^{6}$  & 0.846                       & 3.93                & 349.9                       & $500$                & F \\
\run{m10h}     & $7.95\times 10^{6}$  & 1.45                        & 9.94                & 85.5                        & $500$                & F \\
\run{m10i}     & $8.09\times 10^{6}$  & 1.15                        & 5.68                & 175.7                       & $500$                & F \\
\run{m10j}     & $9.83\times 10^{6}$  & 1.2                         & 3.60                & 58.5                        & $500$                & F \\
\run{m10k}     & $1.06\times 10^{7}$  & 1.25                        & 7.94                & 51.9                        & $500$                & F \\
\run{m10l}     & $1.31\times 10^{7}$  & 1.15                        & 3.77                & 248.8                       & $500$                & F \\
\run{m10m}     & $1.47\times 10^{7}$  & 1.24                        & 6.72                & 106.1                       & $500$                & F \\
\run{m10xe\_D} & $3.8\times 10^{6}$   & 1.04                        & 3.79                & 53.4                        & $4,000$              & G \\
\run{m10xe\_A} & $3.66\times 10^{6}$  & 1.52                        & 3.89                & 53.4                        & $4,000$              & G \\
\run{m10xc\_A} & $8.85\times 10^{6}$  & 0.97                        & 5.60                & 38.3                        & $4,000$              & G \\
\run{m10xe\_B} & $1.33\times 10^{7}$  & 1.24                        & 3.94                & 53.4                        & $4,000$              & G \\
\run{m10xd\_A} & $1.48\times 10^{7}$  & 3.29                        & 2.94                & 78.2                        & $4,000$              & G \\
\run{m10xe\_C} & $2.2\times 10^{7}$   & 1.25                        & 3.62                & 53.4                        & $4,000$              & G \\
\run{m10xg\_A} & $1.96\times 10^{7}$  & 1.89                        & 6.22                & 85.2                        & $4,000$              & G \\
\run{m10xb}    & $3.34\times 10^{7}$  & 2.68                        & 1.94                & 77.6                        & $4,000$              & G \\
\run{m10xh\_A} & $5.48\times 10^{7}$  & 1.77                        & 3.58                & 124.3                       & $4,000$              & G \\
\run{m10xd}    & $7.1\times 10^{7}$   & 4.55                        & 3.06                & 78.2                        & $4,000$              & G \\
\run{m10xa}    & $8.07\times 10^{7}$  & 2.16                        & 6.18                & 343.2                       & $4,000$              & G \\
\run{m10xc}    & $1.21\times 10^{8}$  & 3.93                        & 5.55                & 38.3                        & $4,000$              & G \\
\run{m10xf}    & $1.29\times 10^{8}$  & 6.21                        & 1.17                & 54.6                        & $4,000$              & G \\
\run{m10xe}    & $3.32\times 10^{8}$  & 5.36                        & 4.28                & 53.4                        & $4,000$              & G \\
\run{m10xi}    & $4.32\times 10^{8}$  & 8.79                        & 2.32                & 52.6                        & $4,000$              & G \\
\run{m10xg}    & $4.59\times 10^{8}$  & 7.18                        & 6.01                & 85.2                        & $4,000$              & G \\
\run{m10x}     & $5.23\times 10^{8}$  & 9.19                        & 3.09                & 486.7                       & $4,000$              & G \\
\run{m10q}     & $3.27\times 10^{6}$  & 0.824                       & 6.02                & 95.7                        & $30$                 & H \\
\run{m11h}     & $1.43\times 10^{8}$  & 18.6                        & 4.10                & 146.0                       & $880$                & -- \\
\run{m11b}     & $3.05\times 10^{7}$  & 4.45                        & 2.41                & 144.3                       & $260$                & -- \\
\run{m11q}     & $3.99\times 10^{8}$  & 16.3                        & 3.15                & 31.9                        & $880$                & I \\
\end{tabular}
\caption{Simulations analyzed in this work.  Listed are the names 
of the zoom-in target halo, the stellar mass ($\mstar$) and halo 
mass ($\mhalo$) of that galaxy, the distance to ($\dnearest$) and 
halo mass of ($\mnearest$) the closest other halo with 
$\mhalo>10^{11.5}\msun$, the resolution of each simulation quantified
by the initial baryonic particle mass ($m_i$), and the publication where 
each halo first appeared at the targeted resolution (see citations 
therein for earlier publications that feature lower resolution versions 
of most of the halos).  The references are 
A:  \citet{GK2018}; 
B: \citet{GKDisk}; 
C:  \citet{Wetzel2016}; 
D:  \citet{FIRE2}; 
E:  Samuel et al., in preparation; 
F: \citet{Fitts2017}; 
G:  \citet{Graus2019};
H:  \citet{Wheeler2018}; 
I: \citet{ElBadry2018a}.
}
\label{tab:sims}
\end{table}

These results establish two primary questions, which we explore 
here:  (1) Do the FIRE-2 physics, which accurately reproduce many 
other attributes of the LG dwarf galaxies,  reproduce the observed 
trends with $\mstar$ and environment over a statistical sample of 
dwarf galaxies?  (2)  How do the predicted SFHs vary between 
environments?  For example, do simulations of highly isolated 
dwarf galaxies represent a fair comparison to centrals in the LG, 
and does the presence of a second MW-mass galaxy impact star 
formation in the dwarf galaxies throughout the LG?  In this paper, 
we address these questions by analyzing a large suite of dwarf 
galaxies simulated with identical physics in a variety of 
environments.  We describe the simulations and our sample in 
\S\ref{sec:sims}, present and discuss our results (and caveats 
to those results) in \S\ref{sec:results}, and summarize our 
conclusions in \S\ref{sec:conclusions}.  All of our simulations 
adopt flat $\Lambda$CDM cosmologies with $h \simeq 0.7$ and 
$\Omega_\mathrm{m}\simeq0.3$ \citep[e.g.][]{Larson2011,Planck15,Planck2018}.

\section{Simulations}
\label{sec:sims}

\begin{table*}
    \centering
    \renewcommand{\arraystretch}{1.5}
    \begin{tabular}{lll}
        Environment            & Other terms                                 & Definition    \\
        \hline\hline
        LG satellites          & LG subhalos                                 & within $300~\kpc$ of a MW-mass halo in an LG-like pair  \\ \hline
        LG centrals            & \makecell{Local Field \\ LG non-satellites \\ LG near-field} & more than $300~\kpc$ from all MW-mass halos, but within $2~\mpc$ of a MW-mass halo in an LG-like pair \\ \hline
        isolated MW satellites & subhalos                                    & within $300~\kpc$ of single (non-paired) MW-mass halo  \\ \hline
        isolated MW centrals   & non-satellites                              & between $300~\kpc$ and $2~\mpc$ of a single MW-mass halo \\ \hline
        isolated centrals      & \makecell{dwarf primaries \\ field dwarfs \\ true-field dwarfs}  & no MW-mass halos within at least $2~\mpc$  \\
    \end{tabular}
    \caption{The five environments analyzed in this work.  The second 
    column lists alternative terms for each environment sometimes 
    adopted in the literature.}
    \label{tab:envs}
\end{table*}

All of our dwarf galaxies are taken from cosmological, hydrodynamic 
zoom-in \citep{Katz1993,Onorbe2014} simulations that are a part of 
the FIRE project \citep{FIRE} and run using the ``FIRE-2'' version 
of the code presented in \citet[][i.e. with identical physics 
and code]{FIRE2}.  All simulations were initialized with 
\texttt{MUSIC} \citep{MUSIC} and evolved with \texttt{GIZMO}
\citep{GIZMO}\footnote{\url{http://www.tapir.caltech.edu/~phopkins/Site/GIZMO.html}}
in its meshless finite-mass (``MFM'') mode.  The FIRE physics
modules are described in detail in the papers above; briefly,
we include radiative heating/cooling for $10-10^{10}$~K; allow
for star formation in dense gas that is Jeans unstable, molecular
and self-shielding \citep{Krumholz2011}, and self-gravitating
\citep{Hopkins2013sf_criteria}; and include stellar feedback via
radiation pressure, photo-electric heating and photo-ionization, 
supernovae Types Ia and II and metal mass loss assuming each
star particle is a single stellar population with a \citet{Kroupa2001}
initial mass function.  The simulations adopt the December 2011 
update of the \citet{FaucherGiguere2009} UV background 
model,\footnote{Available at \url{http://galaxies.northwestern.edu/uvb/}.} 
which was designed to produce a reionization optical depth 
consistent with WMAP-7, with a reionization redshift $\zreion\simeq10$.
Most of our simulations (all but those taken from \citealp{Fitts2017}) 
include turbulent metal diffusion \citep{Hopkinsmetaldiff}, 
which yields more realistic stellar metallicity distributions 
in simulated dwarf galaxies \citep{Escala2017} but has a negligible 
effect on the star formation (\citealp{Su2016} and Figure~\ref{fig:m10qres}).

\begin{figure}
    \centering
    \includegraphics[width=\columnwidth]{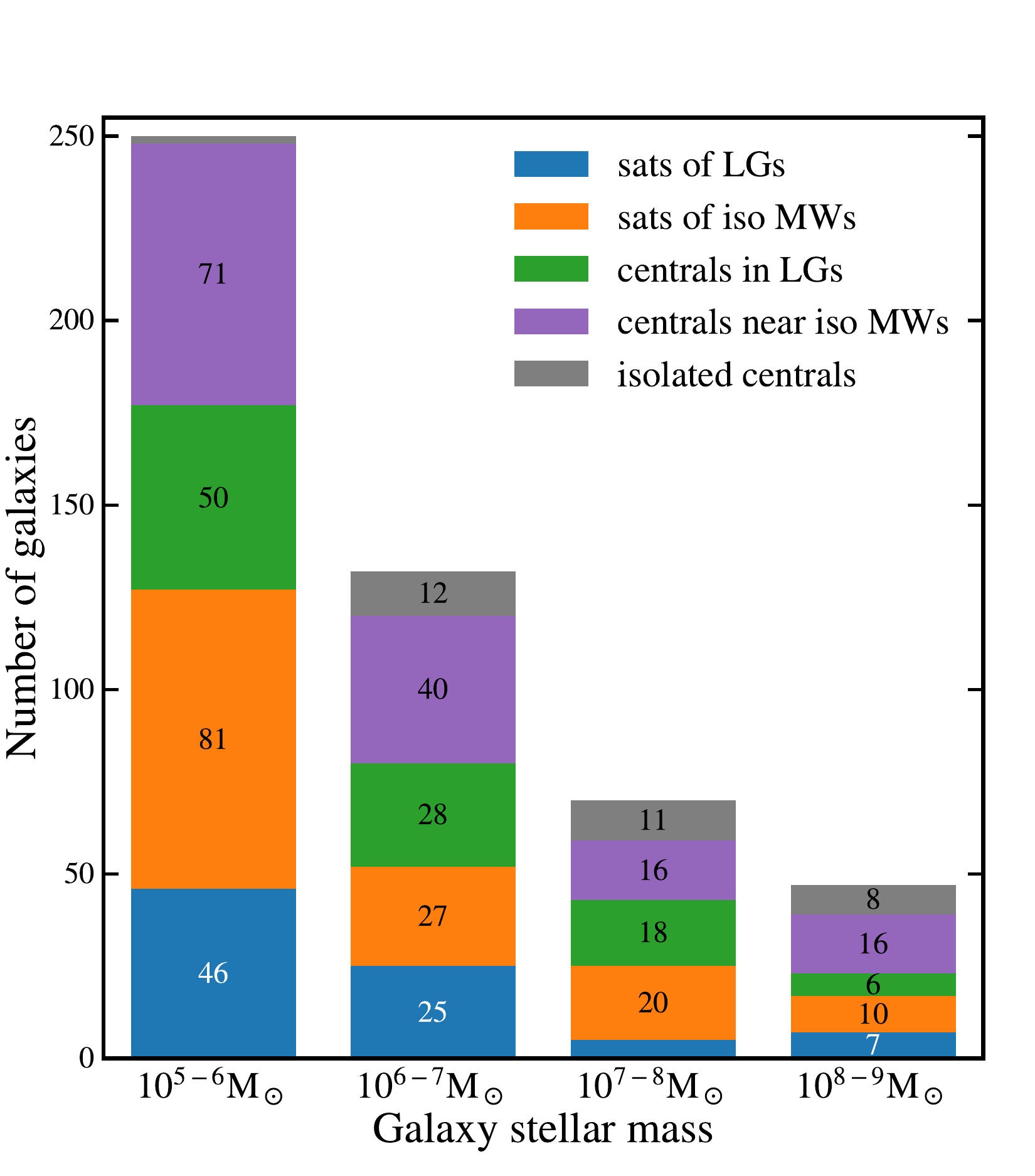}
    \caption{Our simulated sample, split into bins of stellar mass and 
    separated by environment.  It includes truly isolated 
    (``isolated central''; grey) dwarf galaxies that are selected 
    as the targets of a zoom-in simulation along with dwarfs that 
    evolve alongside MW-mass hosts.  We split the latter by their 
    distance to the nearest such host at $z=0$:  satellites are 
    defined as those with $\dhost\leq300~\kpc$ and centrals have
    $\dhost>300$--$2000~\kpc$.  We then further split these samples 
    into ``satellites of LGs'' (blue) and ``centrals in LGs'' (green)~--~taken 
    from the two simulations targeting paired (LG-like) MW-mass 
    hosts~--~and ``satellites of isolated MWs'' (orange) and ``centrals near isolated 
    MWs'' (purple) -- taken from the eight simulations
    targeting individual MW-mass galaxies.  Table~\ref{tab:envs}
    summarizes our environmental definitions.}
    \label{fig:sample}
\end{figure}

\begin{figure}
    \centering
    \includegraphics[width=\columnwidth]{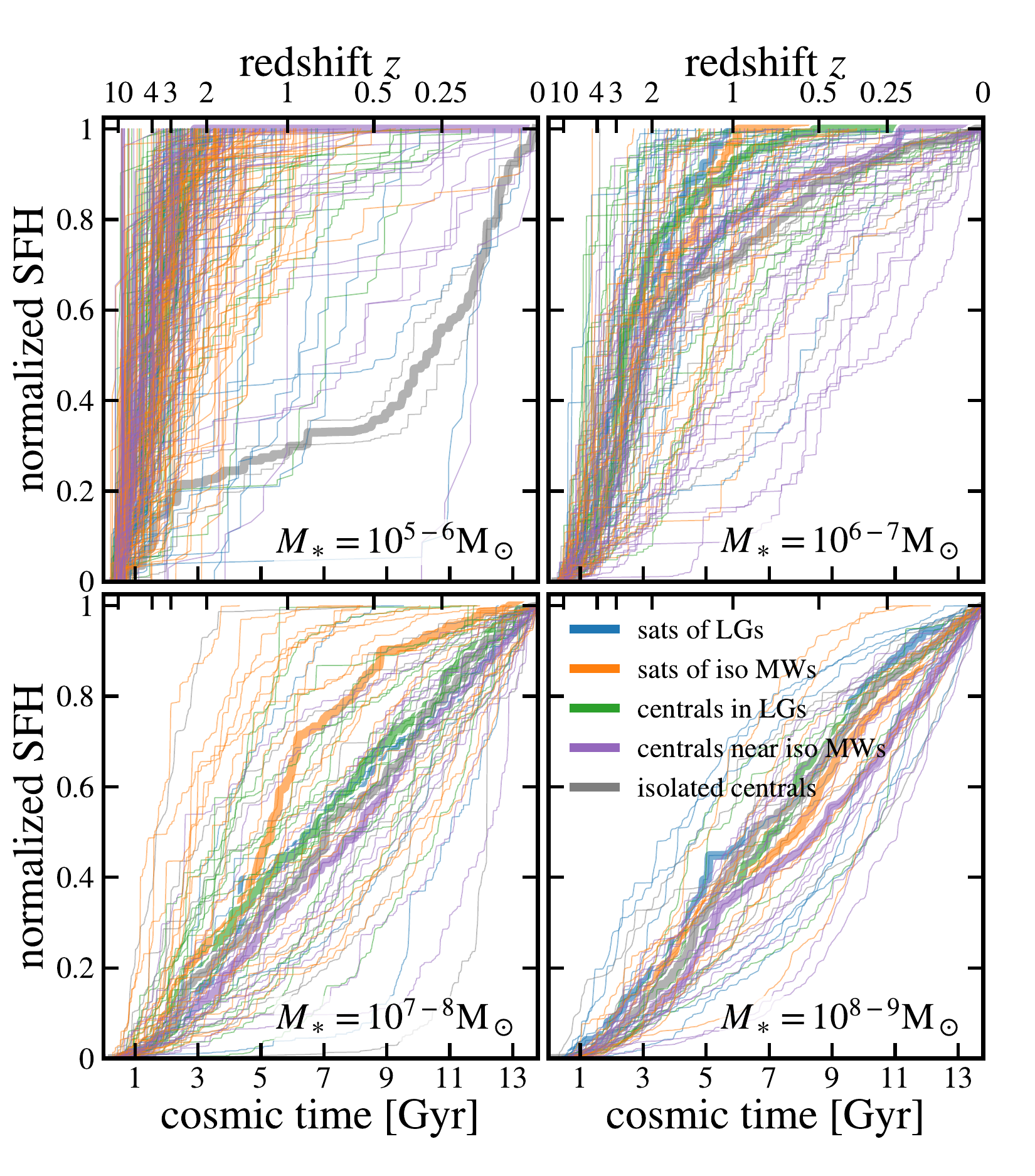}
    \vspace{-2em}
    \caption{Cumulative archaeological star formation histories (SFHs) 
    of all dwarf galaxies in our sample, split into panels of stellar mass 
    at $z=0$ and colored by their environment.  There is a clear trend 
    across environments for higher mass galaxies to form a larger fraction 
    of their stellar mass at later times, though the variety in the 
    detailed SFHs at fixed $\mstar$ is remarkable.  The thick lines 
    in the background plot the medians of each set of dwarf galaxies.  
    As we discuss in \S\ref{ssec:caveats}, our lowest mass dwarf galaxies 
    (with $\mstar\leq10^6\msun$) around MW-mass host(s) may be subject to 
    resolution effects that depress late time star formation.}
    \label{fig:allSFHs}
\end{figure}

\begin{figure*}
    \centering
    \includegraphics[width=\columnwidth]{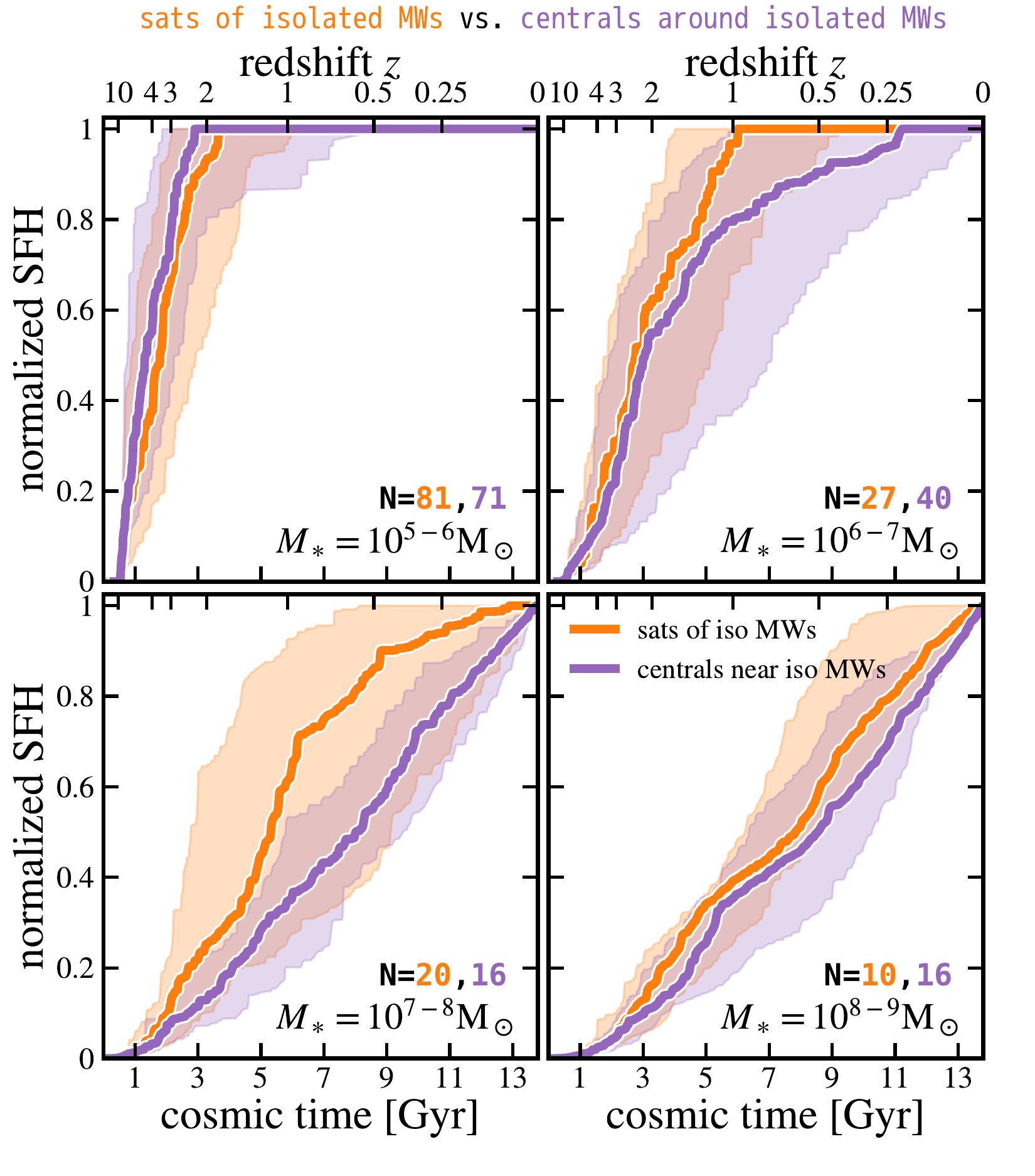} 
    \includegraphics[width=\columnwidth]{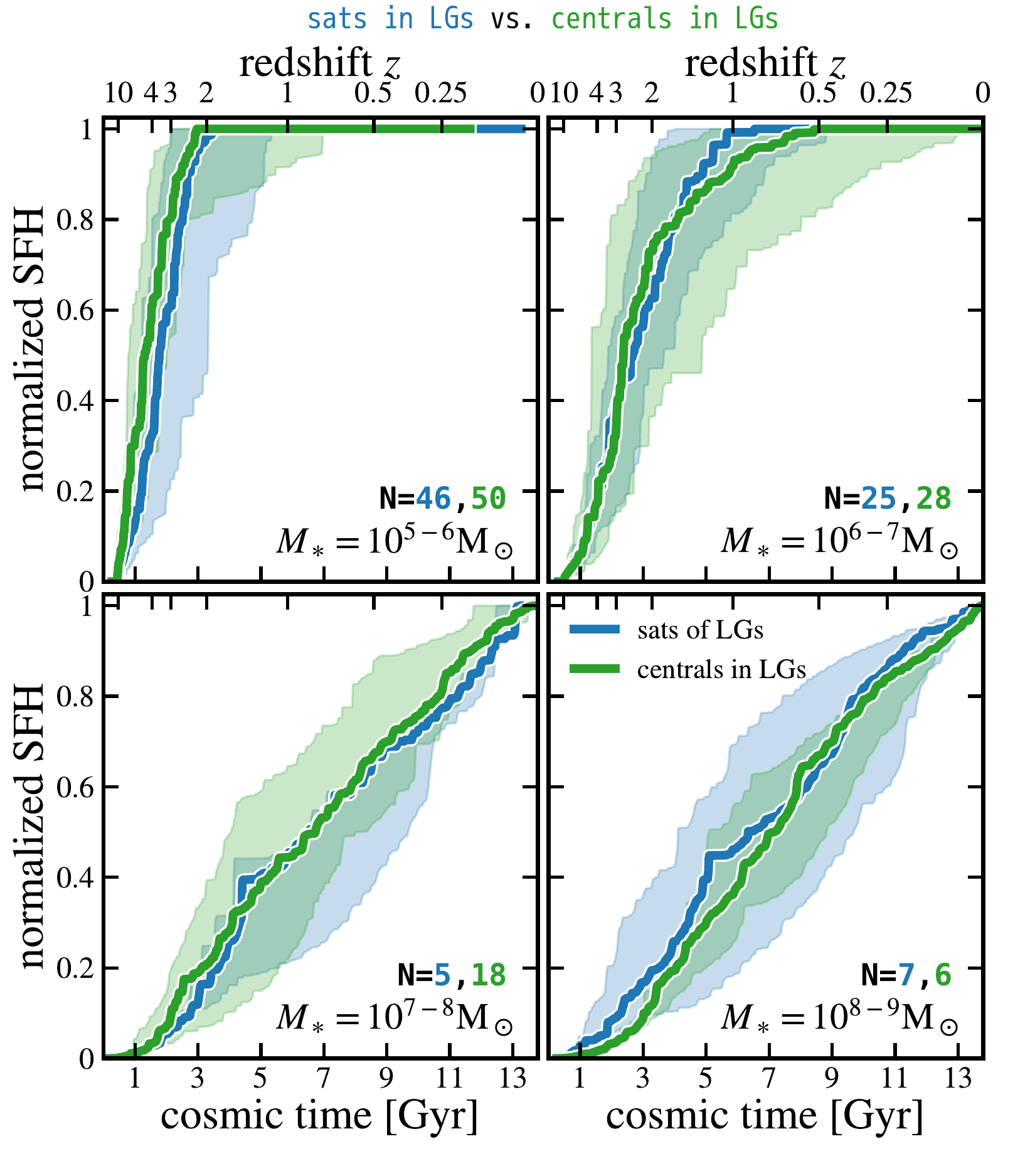}
    \caption{Median SFHs and 68\% scatter for dwarf galaxies
    binned by stellar mass and environment.  The left 
    panel compares satellites ($\dhost\leq300~\kpc$ at $z=0$) 
    to centrals ($\dhost = 300$--$2000~\kpc$) around isolated 
    MW-mass hosts.  Centrals generally have more extended 
    SFHs, consistent with a picture where satellites have their 
    star formation quenched by interactions with the host.  The 
    right panel shows the same comparison with dwarfs from the 
    LG-like simulations.  The satellite/central samples are 
    relatively similar, suggesting that the LG environment 
    impacts star formation in dwarf galaxies even beyond the 
    virial radii of the hosts at a similar level to satellites 
    inside the virial radius.  The numbers in the lower right of 
    each plot give the number of galaxies in each bin.}
    \label{fig:satVfield}
\end{figure*}

We analyze dwarf galaxies from zoom-in simulations that target 
LG-like pairs of MW-mass hosts, isolated MW-mass hosts, and 
highly isolated dwarf central galaxies (i.e. without including
any MW-mass hosts in the zoom-in volume).  We plot our sample 
as a function of stellar mass, using decade-wide bins, and 
separated by environment in Figure~\ref{fig:sample}, and list
the parent simulations for the full sample in Table~\ref{tab:sims}, 
which gives the primary galaxy (or, in the case of the LG-like 
runs, galaxies) in each run along with their stellar and halo 
masses, a measure of their isolation, mass resolution, and the 
publication where each halo first appeared at the adopted 
resolution.\footnote{
    Our sample includes two new isolated MW-mass hosts, \run{m12r} 
    and \run{m12w}, selected to contain Large Magellanic Cloud-like 
    satellites in their dark matter-only parent simulations, though 
    neither contains such a satellite in the zoom-in runs we analyze 
    here.  Both hosts will be presented in greater detail in Samuel 
    et al., in preparation.
}
We also direct the reader to \citet{ElBadry2018a,ElBadry2018}, who 
analyzed the HI properties of the majority of the isolated central 
sample; to \citet{m12morph}, who analyzed the morphologies and growth 
histories of the majority of the MW-mass hosts (though typically at 
lower resolution); and to \citet{Sanderson2017}, who studied the mass 
in the stellar halos of the MW-mass hosts (again typically at lower 
resolution).  

In the LG and isolated MW simulations, we identify (sub)structure in 
the dark matter particles using the \texttt{Rockstar} \citep{rockstar} 
6D halo finder, then use a similar (though not identical) process as 
\citet{Necib2018} to assign star particles to those overdensities. 
The method is described in detail in Samuel et al. (in preparation)
but, in short, we select star particles that are located within the 
radius of the halo (as reported by \texttt{Rockstar}) and moving with
a relative velocity that is within $2\times\vmax$.  We then define 
$R_{90}$ as the radius that encloses 90\% of that stellar mass and 
$\sigma_\ast$ as the velocity dispersion of those star particles.  
Finally, we iteratively remove stars that are $>1.5\times R_{90}$ from 
either the galaxy or (sub)halo center, or that have a velocity offset
$>2\times\sigma_\ast$, until the stellar mass converges to within 1\%.\footnote{
    The catalogs therefore differ from those used in \citealt{GK2018}; 
    most notably, those were based on halos identified by the spherical 
    overdensity-based \texttt{AHF} \citep{AHF}.  We have confirmed 
    that the stellar mass functions and circular velocity profiles 
    obtained via the new \texttt{Rockstar} catalogs are consistent 
    with those of \texttt{AHF}.
}
We find that this method accurately and reliably separates real 
galaxies from transient alignments between subhalos and stars in
the stellar halos of the MW-mass hosts.  Meanwhile, for the highly 
isolated dwarf central galaxies, which do not overlap with the extended
stellar halos that surround MW-mass hosts, we adopt all particles that are within the radius 
that contains 90\% of the stellar mass within $20~\kpc$ of the 
galaxy center.  In all cases, we define $\mstar$ as the sum of the 
masses of the member star particles and calculate SFHs using their 
formation times.  Our smallest galaxies (with $\mstar\simeq10^5\msun$) 
in our lowest resolution simulations (the isolated MW ``Latte'' runs, 
with initial gas particle masses $m_i = 7,100\msun$) therefore 
contain a minimum of $14$ star particles, though stellar mass 
loss reduces the mass per particle such that the smallest galaxy
we analyze actually contains $\sim20$ star particles.  We discuss the 
potential for resolution artifacts, and their impact on our 
conclusions, in \S\ref{ssec:caveats} and Appendix~\ref{sec:resolution}.

\begin{figure*}
    \centering
    \includegraphics[width=\columnwidth]{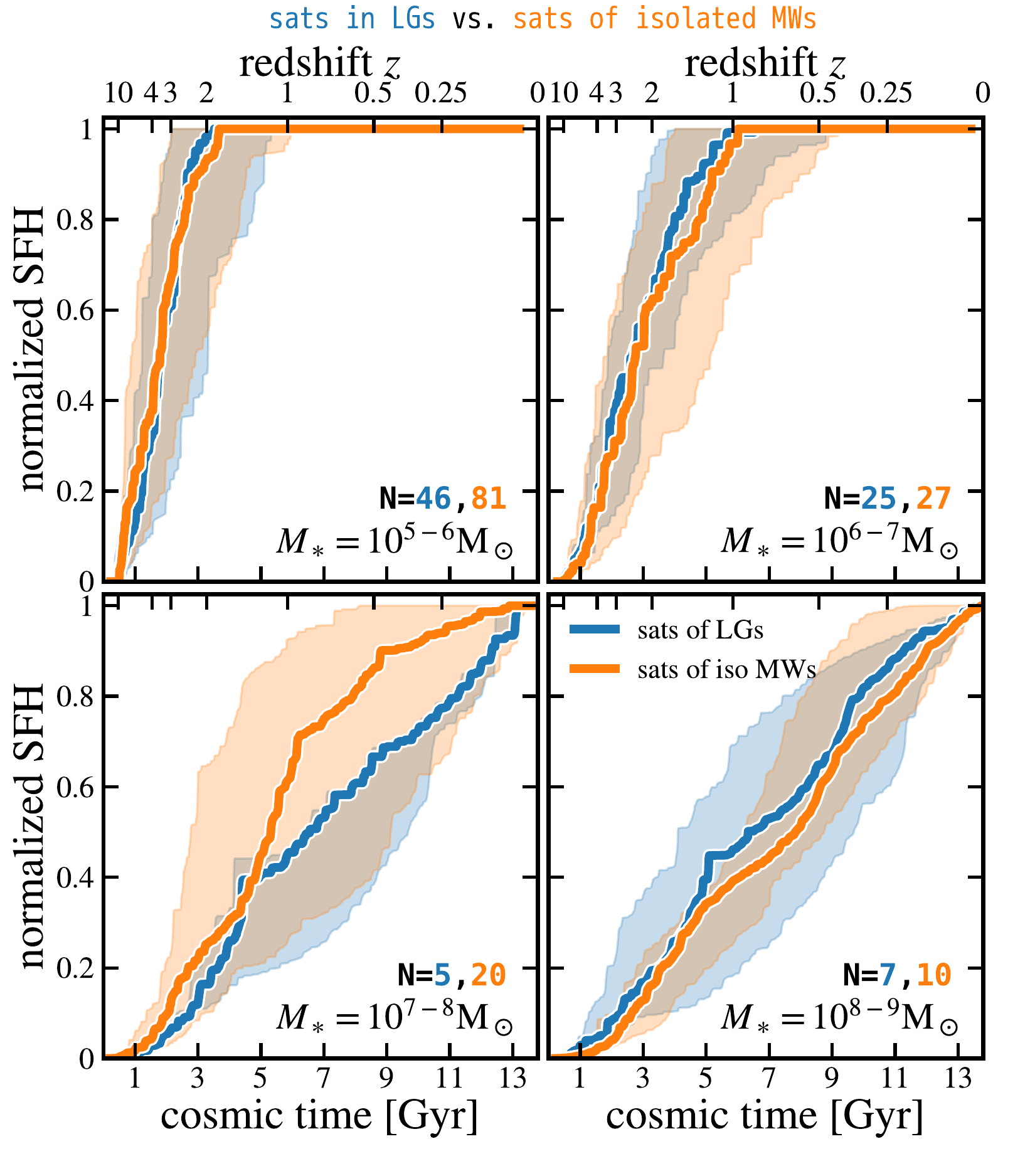}
    \includegraphics[width=\columnwidth]{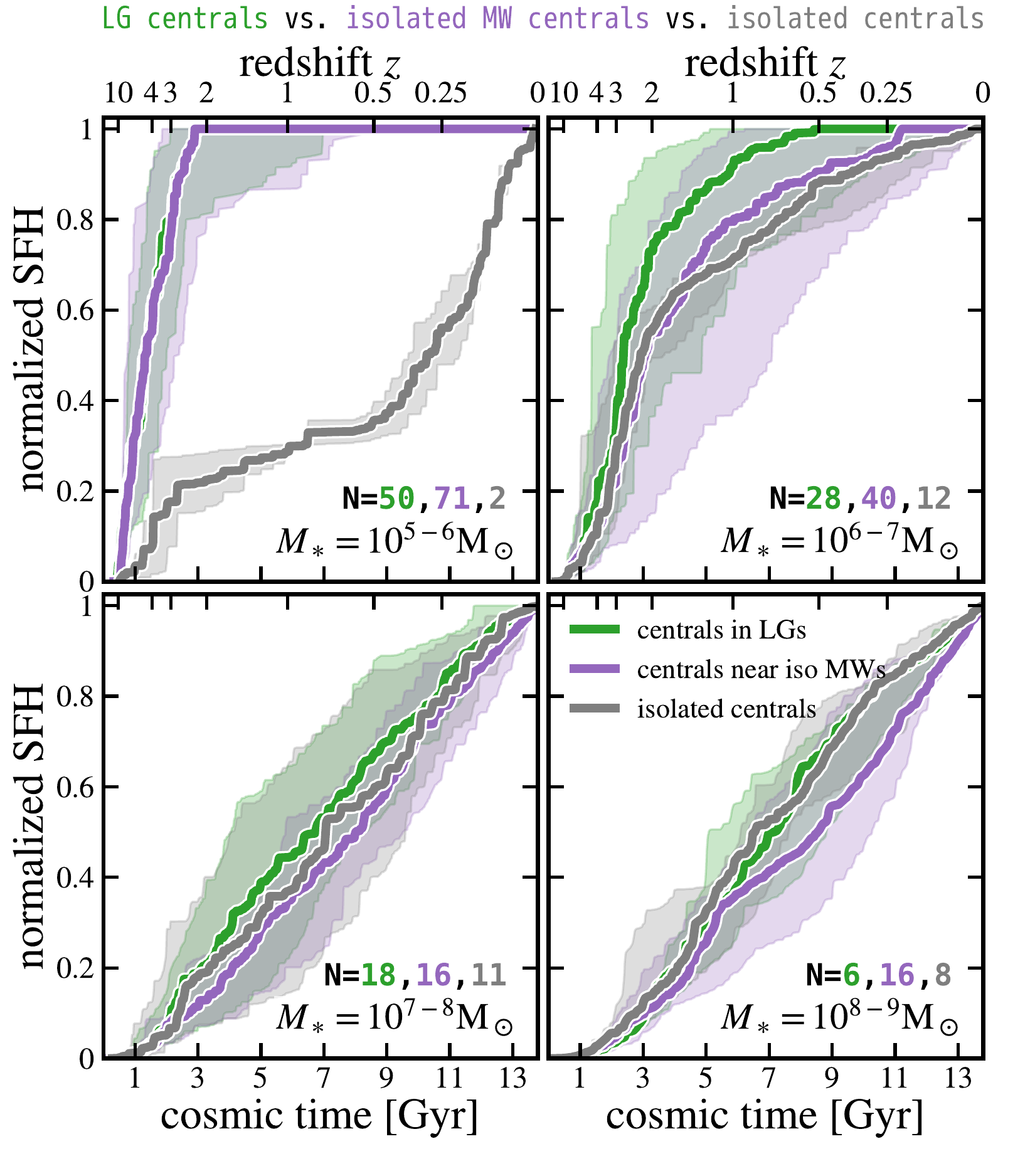} 
    \caption{The SFHs of the satellites (left) and centrals 
    (right) around isolated MWs compared to those in LG-like 
    pairs and (in the right panel) those far from any MW-mass 
    host.  The lines and shaded regions again indicate the 
    median and 68\% contours. The satellite populations 
    are similar at nearly all masses, though those in LG-like 
    environments do tend to form marginally earlier at fixed $\mstar$.  
    The lone exception is for $\mstar=10^7$--$10^8\msun$, but 
    the LG satellite sample includes only five galaxies in that 
    mass range.  However, the SFHs of dwarf centrals exhibit 
    stronger environmental trends:  those around isolated MWs 
    tend to form later than LG centrals at roughly all masses, 
    and particularly for $\mstar=10^6-10^7\msun$.  Though the 
    shift in the medians is within the scatter, we note that 
    the scatter also tends to shift in the same 
    sense as the median.  For $\mstar\leq10^7\msun$, the highly 
    isolated dwarf centrals have SFHs that are even more extended 
    than the centrals around isolated MWs.  Isolated centrals with
    $\mstar=10^5$--$10^6\msun$ display highly discrepant SFHs compared
    to dwarf centrals with at least one MW-mass host nearby, but we
    caution that the latter sample may be affected by resolution 
    (see \S\ref{ssec:caveats}) and that there are only two galaxies 
    in the former sample.}
    \label{fig:LGvIso}
\end{figure*}

Throughout, we take dwarfs within $300~\kpc$ of a MW-mass host at 
$z=0$ as ``satellites'' while dwarf galaxies more than $300~\kpc$ 
from a MW-mass galaxy are classified as ``centrals.''\footnote{%
    Our qualitative conclusions are insensitive to this choice:  the
    median shapes change only slightly as we vary the satellite/central 
    cut from $250$~--~$400~\kpc$.}
We further separate satellites into those of isolated MWs and of 
hosts in LG-like pairs, and split the centrals into those around 
isolated MWs, those in an LG-like environment, and highly isolated 
dwarf centrals that are the primary target of their zoom-in volumes.  
Table~\ref{tab:envs} summarizes the different environmental
definitions and presents alternative names sometimes adopted 
in the literature.

For consistency with observations, we present archaeological SFHs 
throughout, calculated by taking the time that each star in the 
galaxy at $z=0$ formed.  Therefore, some fraction of the stars 
included in the SFHs may have formed in an external galaxy and 
been brought in via mergers. Using the same \citet{Fitts2017} 
sample of isolated dwarf centrals as adopted here, \citet{Fitts2018} 
found that this fraction is typically small ($<10\%$) for 
$\mstar\lesssim10^7\msun$, and \citet{AnglesAcazar2017} found
a similarly small fraction in a simulated FIRE galaxy with 
$\mstar=1.4\times10^9\msun$.  While \citet{Deason2014} used 
dark matter-only zoom-in simulations (both of isolated MW-mass 
halos and of LG-like pairs) to show that most dwarf halos in LG-like environments have 
undergone a major merger at some point in their evolution 
(roughly $45-70\%$, with mergers more common among centrals 
and higher mass dwarf halos) the vast majority of those mergers 
occur at cosmic time $t\lesssim3~\gyr$, before the majority of 
star formation in most of our sample \citep[and also see][]{RodriguezGomez2016}.

\section{Results}
\label{sec:results}

Figure~\ref{fig:allSFHs} summarizes the SFHs.  Each panel plots 
the star formation histories (SFHs) of the galaxies in our sample 
within a given decade of galaxy mass.  The thin lines plot the 
individual galaxies, while the thick lines take the median of
each environment.  

Figure~\ref{fig:allSFHs} reveals two conclusions, which generally 
agree with previous results from both simulations and observations.
First, for $\mstar\lesssim10^8\msun$, there is an obvious trend with 
galaxy mass (across environments) where higher mass galaxies form a 
higher fraction of their stars at later times.  Galaxies with 
$\mstar=10^{5-6}\msun$, which appear to often be dominated by starvation 
following reionization, typically quench (stop forming stars) by 
$t\sim3~\gyr$, while galaxies with $\mstar\geq10^8\msun$ almost 
universally continue to form stars to $z=0$.  Second, even within a 
fixed mass bin (and in a fixed environment), there is a large degree 
of scatter in the SFHs.  Even with this scatter, though, we find it 
exceptionally rare for galaxies to form their first stars at 
$t\gtrsim 1~\gyr$ ($z\lesssim6$).  Only a few such galaxies exist
in our sample, nearly all with $\mstar\leq10^6\msun$.  Moreover,
only one galaxy (a low-mass central around an isolated MW-mass 
host) forms its first star after $t\simeq4~\gyr$, suggesting that
galaxies of this type are indeed very rare in \lcdm, in contrast
with the predictions of warm DM models \citep{Bozek2018}.

\subsection{Environmental variations}
We now turn to the impact of environment on the shape of dwarf 
SFHs.  We focus on comparing ``satellite \emph{vs.} central'' 
and ``isolated MW \emph{vs.} LG-like pair \emph{vs.} isolated 
dwarf central,''  but we will present statistics for all possible 
pairings below.

\subsubsection{Satellites \emph{vs.} Centrals}
We begin with Figure~\ref{fig:satVfield}, which compares the 
SFHs of satellite galaxies to central galaxies.  The left panel 
selects only those objects that evolve around a single, isolated 
MW-mass host.  As expected in environmental quenching models, 
satellites tend to reach a given fraction of their $z=0$ stellar mass 
at earlier times than centrals of a similar final mass.  This is particularly 
evident for $\mstar=10^6-10^8\msun$, but the tail of late time star 
formation in centrals is longer than that of the satellites even for 
$\mstar=10^5-10^6\msun$.

In contrast, the right panel of Figure~\ref{fig:satVfield} demonstrates 
that if such a distinction exists in the LG-like environments, it is
strongly muted:  satellites and centrals display far more similar
behavior overall with a smaller tail of late-time star formation 
for centrals with $\mstar\leq10^7\msun$ than in the environments 
of isolated MWs.

\subsubsection{Local Groups \emph{vs.} isolated MWs \emph{vs.} the field}

Figure~\ref{fig:LGvIso} therefore compares the LG, isolated MW,
and isolated dwarf samples, separating satellites and centrals.  
The median SFHs of the satellite populations (left panel) of the 
LGs and isolated MWs are reasonably similar at most masses, though 
the LG satellites tend to form marginally earlier.  The only deviation 
from this trend, and the mass range where the LG satellites differ 
the most from their isolated-MW counterparts, is for 
$\mstar=10^7-10^8\msun$.  However, we caution that the LG-like 
simulations contain only five satellites in that mass range.

The right panel of Figure~\ref{fig:LGvIso} now compares the
SFHs of dwarf central galaxies, including those around isolated 
MWs, those around LGs, and ``true field'' dwarf centrals with 
no nearby MW-mass host.  As expected from Figure~\ref{fig:satVfield}, 
the former two samples exhibit clear differences, particularly for 
$\mstar=10^6-10^8\msun$, where dwarf centrals in LGs tend to form 
their stars earlier than their counterparts around isolated MW-mass 
hosts.  The offset is largest for $\mstar=10^6$--$10^7\msun$, where 
the medians are offset by nearly the full 68\% contours.  We also 
emphasize that both the medians and the scatters tend to shift relative 
to one another in the same sense.  We examine the statistical 
significance of this result in \S\ref{ssec:summary_stats} and 
argue in \S\ref{ssec:caveats} that it is robust to resolution.

\begin{figure}
    \centering
    \includegraphics[width=\columnwidth]{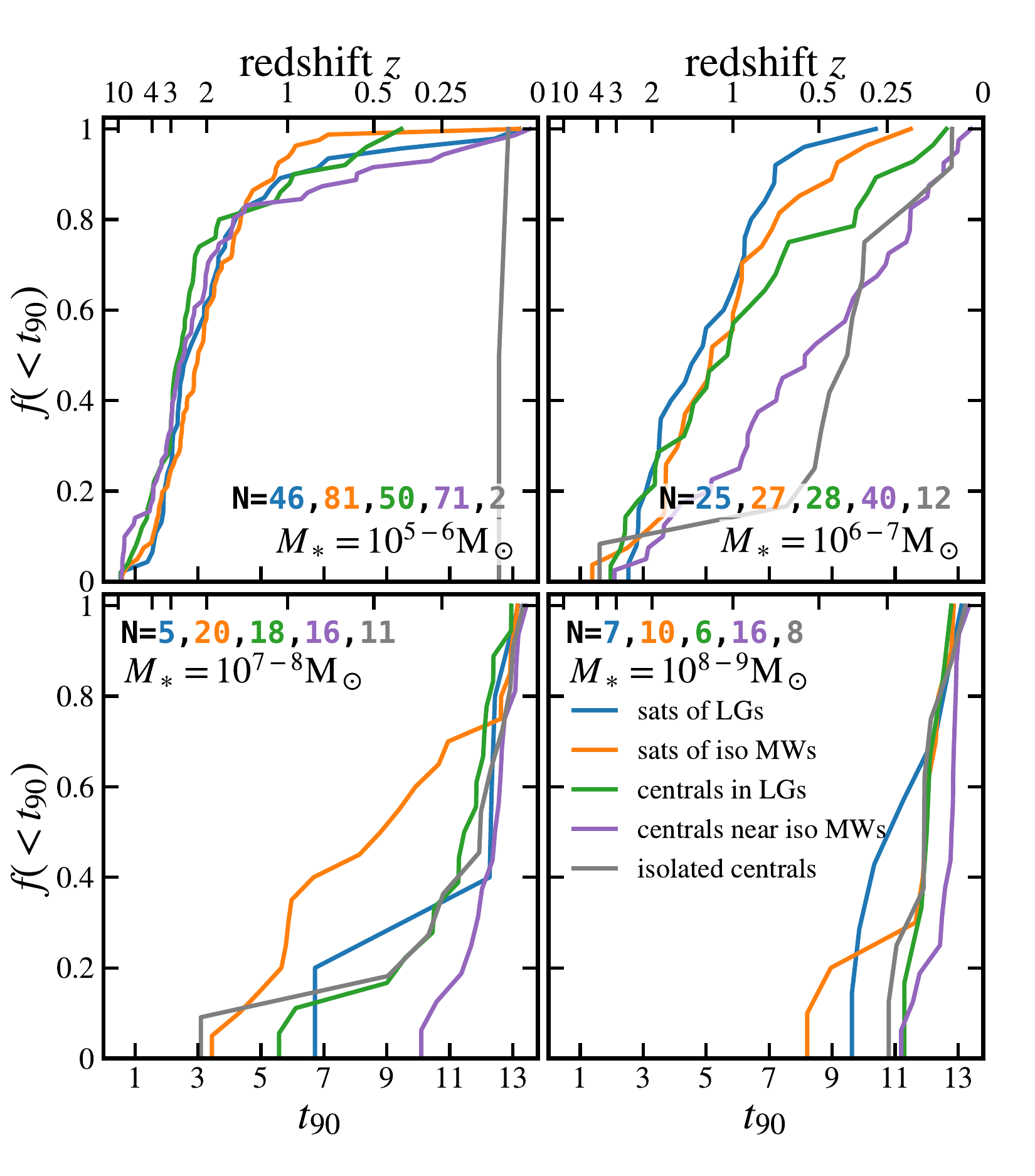} 
    \vspace{-2em}
    \caption{Distributions of the approximate quenching times of dwarf galaxies, 
    quantified by the cosmic time when each galaxy reached 90\% ($\tnintey$) of 
    the stars in it at $z=0$.  The distributions are computed by binning the 
    dwarfs by stellar mass (panels) and environment (line color; see legend).  
    There is a strong trend for more massive dwarfs to quench later (if at 
    all), regardless of environment.  Within a given mass bin, satellite 
    distributions are typically shifted to earlier times relative to centrals.
    As expected from previous Figures, however, the differences between LG
    centrals and satellites is much smaller than that between centrals around isolated
    MWs and satellites.  Figure~\ref{fig:ADmatrix} quantifies the significance
    of these differences for $\mstar=10^6$--$10^7\msun$.
    }
    \label{fig:moments}
\end{figure}

\begin{figure}
    \centering
    \includegraphics[width=\columnwidth]{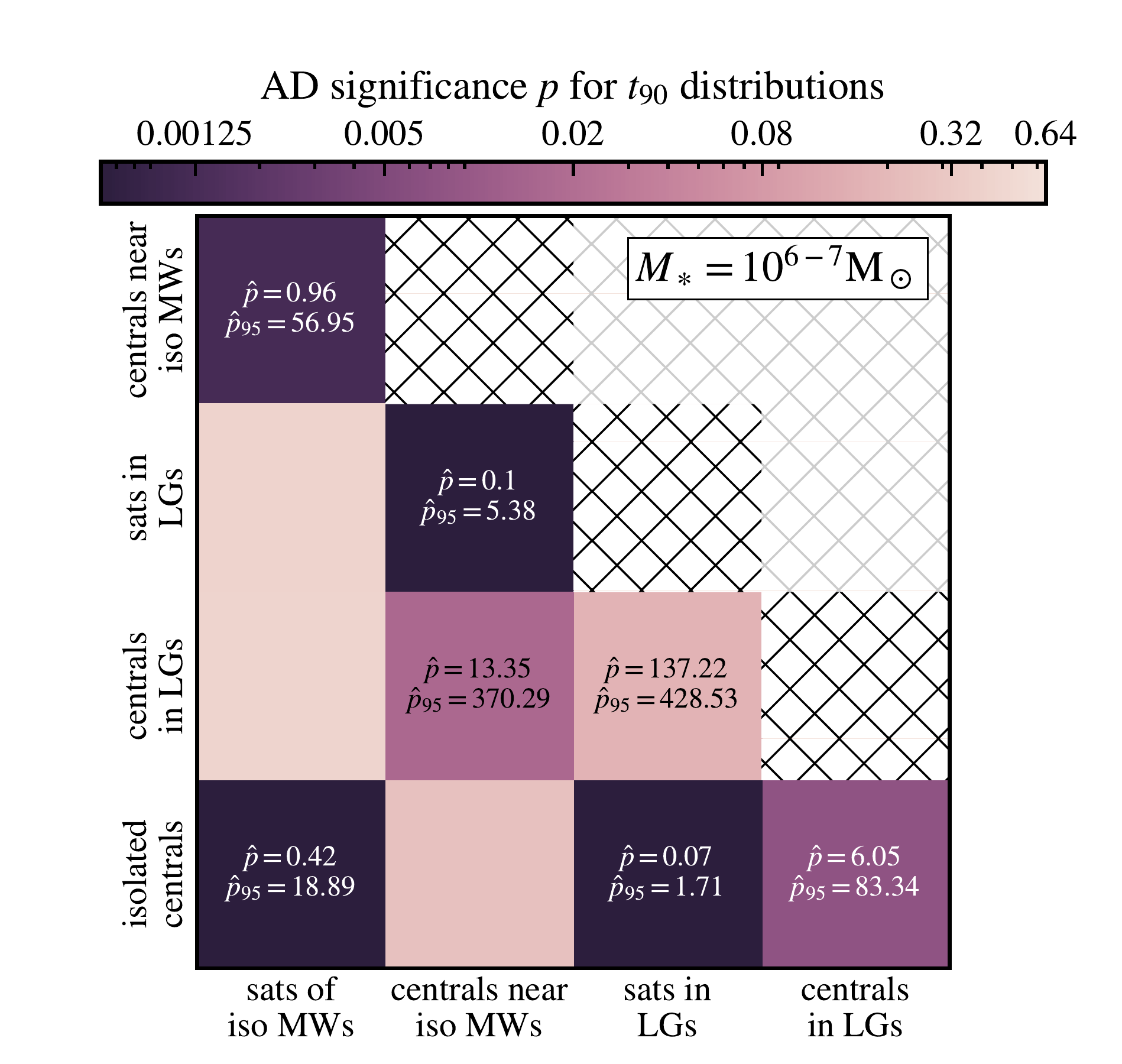} 
    \vspace{-2em}
    \caption{The statistical significance of the differences in the $\tnintey$ 
    distributions (in the $10^6$--$10^7\msun$ bin) between any pair of 
    environments, quantified by the Anderson-Darling statistic (see 
    footnote~\ref{footnote:ad-desc}).  $\hat{p}\leq1$ indicates a 
    statistically significant difference between the samples on the
    $x$ and $y$ axes.  Though it captures only a portion of the 
    differences in the SFHs, the AD test shows that satellites of 
    isolated MWs, LG satellites, and LG centrals, do \emph{not} differ 
    significantly from one another.  However, all three of these 
    populations differ from (quench earlier than) isolated dwarf centrals
    (bottom row) and centrals around isolated MWs.  The difference between 
    LG centrals and the latter two populations does not pass our corrected
    significance threshold, but the lack of a difference between satellites
    and centrals in LGs (especially compared to the significant offset between 
    satellites and centrals around isolated MWs) is itself noteworthy.
    We discuss the results of the other mass bins and the other statistics
    in the text.
    }
    \label{fig:ADmatrix}
\end{figure}

The truly isolated dwarf centrals appear to continue the trend
between the number of nearby MW-mass hosts and the fraction of 
late-time star formation for $\mstar\leq10^7\mstar$:  at fixed 
stellar mass, the isolated dwarf centrals tend to form later
than those with MW-mass neighbors.  While the difference in the
lowest mass bin may be driven (at least partially) by resolution
and our small sample size, the offset persists for 
$\mstar=10^6$--$10^7\msun$.  Though the difference compared to
centrals around isolated MWs is small, it is significant when 
compared with the centrals in LG-like environments.  Therefore, 
the FIRE simulations predict that isolated dwarf galaxies cannot
necessarily be fairly compared with dwarfs within $\sim2~\mpc$
of the LG~--~even those that are dwarf centrals~--~as the former
will have formed more of their stars at late times.

\subsubsection{Summary statistics of the SFHs}
\label{ssec:summary_stats}
To quantify these trends, we summarize the shapes of the SFHs via 
$\tten$, $\tfifty$, and $\tnintey$, the cosmic times when each galaxy 
reaches 10\%, 50\%, and 90\% of its $z=0$ stellar mass, respectively.
Though we do not plot them, both $\tten$ and $\tfifty$ increase with 
$\mstar$ (i.e. more massive dwarf galaxies have more extended SFHs), 
but do not display any clear systematic environmental variations within 
a given mass bin, particularly at low masses where the sample sizes are
large.

The $\tnintey$ distributions, which are plotted in Figure~\ref{fig:moments}, 
display similar mass trends, but also exhibit clear environmental variations,
consistent with our previous conclusions.  As expected, these differences
are strongest for $\mstar=10^6$--$10^7\msun$, but centrals around isolated
MWs (purple lines) tend to have slightly later $\tnintey$ than either the 
LG centrals or the satellite populations at all masses.  Taken together 
with the lack of a difference in the $\tten$ and $\tfifty$ distributions,
Figure~\ref{fig:moments} suggests that the main differences between the 
SFHs of dwarfs in different environments are in the amount of late-time 
star formation that occurs.

We quantify the statistical significance of the differences between 
the distributions of $\tnintey$ via the Anderson-Darling (AD) test 
\citep{Anderson1956:ADTest} in Figure~\ref{fig:ADmatrix}.\footnote{
    The AD test is an improved version of the well-known Kolmogorov-Smirnov 
    (KS) test; we specifically adopt the $k$-sample generalization from 
    \citet[][and also see \citealp{Scholz1987:kSampADTest}]{Pettitt1976:2SampADTest}.  
    The AD significance $p$ is the probability of finding differences in the 
    distributions \emph{at least} as extreme as those observed if the null 
    hypothesis~--~that the two samples are drawn from the same underlying 
    distribution~--~is correct.  A low $p$-value therefore indicates a low probability 
    of finding the observed differences, and therefore renders the null 
    hypothesis unlikely.  We chose a significance threshold that accounts 
    for false positives by adopting the Bonferroni correction, which divides 
    the nominal significance threshold $p\leq0.05$ by the number of comparisons 
    performed \citep{Dunn1961}.  We therefore consider cases with $p\leq0.00125$ 
    as statistically significant and show the normalized $p$-value, 
    $\hat{p} = p/0.00125$, such that $\hat{p} \leq 1$ indicates a statistically 
    significant difference between two samples.  We only show values for $p < 0.25$ 
    ($\hat{p} < 200$).  We also show $\hat{p}$ corresponding to the upper 95\% 
    contour obtained by repeating the AD test on bootstrapped versions of the 
    distributions, which we define as $\hat{p}_{95}$.
    \label{footnote:ad-desc}
}
We highlight the $\mstar=10^6$--$10^7\msun$ bin, which Figure~\ref{fig:moments} 
shows has the largest environmental variations.  Figure~\ref{fig:ADmatrix} confirms 
that several of the differences in the $\tnintey$ distributions are statistically 
significant:  in particular, both satellite populations quench earlier than both 
the dwarf centrals around single MWs and the highly isolated dwarf centrals.  LG 
dwarf centrals, meanwhile, sit in between the two extremes, but, from Figure~\ref{fig:moments}, far more closely 
resemble the satellite populations.

Though we do not plot them, we summarize the other mass bins here:  
for $\mstar=10^5$--$10^6\msun$, all of the $\tnintey$ distributions 
except that of the isolated dwarf centrals are broadly consistent 
with one another.  The latter quenches much later, but our small 
sample of isolated dwarfs with $\mstar=10^5$--$10^6\msun$ renders 
only the difference compared to LG centrals truly significant.  
There are no statistically significant offsets at higher masses, 
where larger samples are needed, but there are hints of differences
between the LG centrals and isolated-MW centrals ($\hat{p}=12.7$ 
and $49.9$ for $\mstar=10^{7-8}$ and $10^{8-9}\msun$, respectively) 
and between the satellites and centrals around isolated MWs 
($\hat{p}=2.6$ and $11.8$ for the same bins).

We also do not plot the results of the AD test on the distributions 
of $\tten$ and $\tfifty$, which confirm that these statistics are
largely insensitive to environment.  The only significant variations
arise in our lowest mass bin with the $\tfifty$ times of isolated 
dwarf centrals, which reach their half-mass times much later than 
the other four environments.  However, many of the $\tfifty$ 
comparisons in this same mass bin yield $\hat{p}$ values that 
suggest changes with environment:  only the ``LG satellite \emph{vs.}
isolated-MW satellite'' and ``LG central \emph{vs.} isolated-MW
central'' comparisons yield $\hat{p}\gtrsim7$ for
$\mstar=10^5$--$10^6\msun$.

\subsubsection{Why do dwarf centrals in LGs form their stars earlier?}
\label{ssec:why_lgs_early}

As the preceding sections showed, the SFHs of dwarf satellites
in our sample appear independent of the larger scale environment 
(i.e. whether or not the host is in a LG-like pair), but the SFHs 
of our dwarf centrals in LGs exhibit less late time star formation 
(at fixed final mass) than their counterparts around isolated MWs.  
In this section, we discuss possible explanations for the offset 
between the centrals~--~and the lack of an offset between the 
satellites~--~across the two environments.

\begin{figure}
    \centering
    \includegraphics[width=\columnwidth]{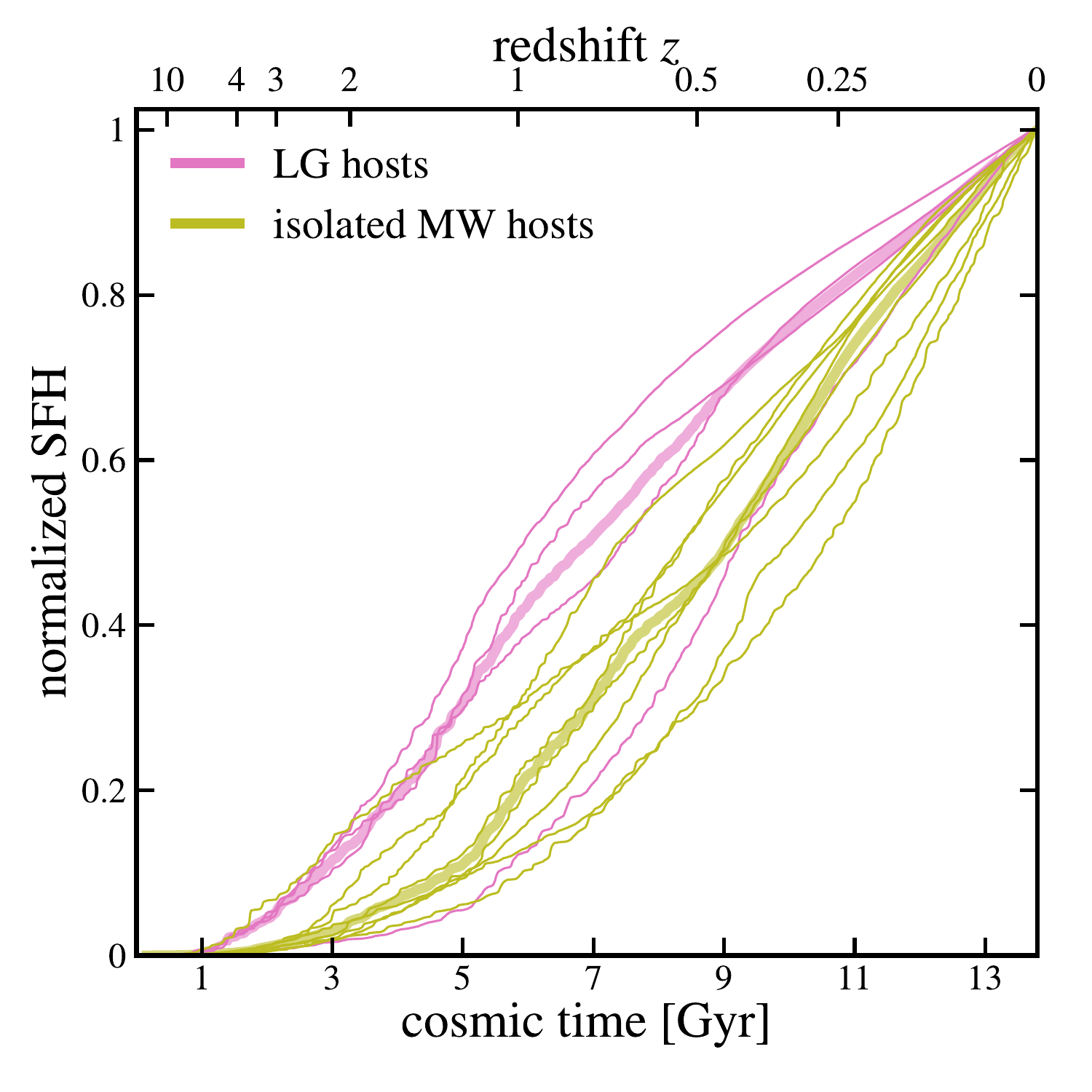}
    \vspace{-2em}
    \caption{The archaeological SFHs of the MW-mass hosts in 
    the simulations, separated into paired and isolated MWs.
    Three of the four paired hosts form half of their stars 
    before any of the isolated MWs reach that milestone.}
    \label{fig:hostSFHs}
\end{figure}

We begin by exploring whether the offset in the SFHs can be explained
via the dark matter accretion histories of the dwarf galaxies, which we
quantify with $\vmax(t)$.  Though it is much smaller than the offset
between the SFHs, we do find that the dwarf centrals in the LGs and
around isolated MWs display slightly different behaviors:  while the
median $\vmax(t)$ curve for centrals in LGs is slightly falling at 
late times, the corresponding curve in the isolated-MW runs is flat 
or slightly rising until $z=0$, particularly for the dwarf halos 
hosting galaxies with $\mstar\leq10^7\msun$.  While the difference
manifests at later times than the offset between the SFHs 
(e.g. for $\mstar=10^{6-7}\msun$, the median $\vmax(t)$ curves cross 
at $t\sim7~\gyr$ while the median SFHs diverge at $t\sim2~\gyr$), 
the trend is in the direction expected if gravitational interactions 
are responsible for the offset in the SFHs.  These interactions could 
be direct, e.g. in the form of an increased fraction of ``backsplash'' 
halos in LGs (centrals that were previously satellites; 
\citealp{Teyssier2012,ELVIS}), or indirect, e.g. if the overall 
LG gravitational potential inhibits late time accretion or if 
structure as a whole assembles earlier in LG environments.  
However, while the median $\vmax(t)$ for satellite galaxies 
is falling at late times in both environments (as expected), 
it falls slightly faster around the LG-like hosts; this is not 
reflected in the SFHs, which are roughly identical.  Moreover, 
the change in $\vmax(t)$ moving from centrals to satellites (in 
either environment) is much larger than when comparing isolated-MW satellites (centrals) to LG satellites (centrals).  We also find 
no evidence of an increased backsplash fraction in the distributions 
of pericentric distances, which do not differ systematically between 
the LG and isolated-MW environments. Therefore, while this explanation
is qualitatively consistent and suggests that the difference may be 
due to a dynamical process, the fact that the trend is not reflected 
in the satellites and the relatively small size of the effect 
prohibits strong statements that the whole of the offset can be 
attributed to the dark matter accretion history or gravitational 
influences.

The offset in the SFHs of centrals could also be tied to the distribution of present day distances from the nearest MW-mass 
host.  The offsets between satellites and centrals suggest that
quenching typically occurs later further from the host; therefore, 
if our LG centrals are typically further from their nearest host 
at $z=0$ than dwarf centrals around isolated MWs, the former may 
appear to quench earlier.  We do find evidence, via $\tnintey$
\emph{vs.} $z=0$ distance, that quenching occurs (on average) 
later for galaxies further from a host today:  fits to the data 
are nearly always rising with distance.\footnote{%
    The lone exception is in the lowest mass bin, where the fit 
    to the LG dwarfs is very effectively flat.
}%
However, dwarfs in LG-like environments quench (on average) earlier 
than their isolated-MW counterparts at roughly all $z=0$ distances. 
The difference is well within the scatter in all the mass bins, but it
is most pronounced for $\mstar=10^6$--$10^7\msun$.  Therefore, the
offset cannot be attributed to the distribution of $z=0$ distances.
Moreover, we find no clear evidence of the difference between LG 
dwarf centrals and dwarf centrals around isolated MWs disappearing 
at large distances.

Why, then, do the LG dwarfs (in both stars and DM) finish their 
formation or ``growth'' phases earlier?

\begin{figure*}
    \centering
    \includegraphics[width=\columnwidth]{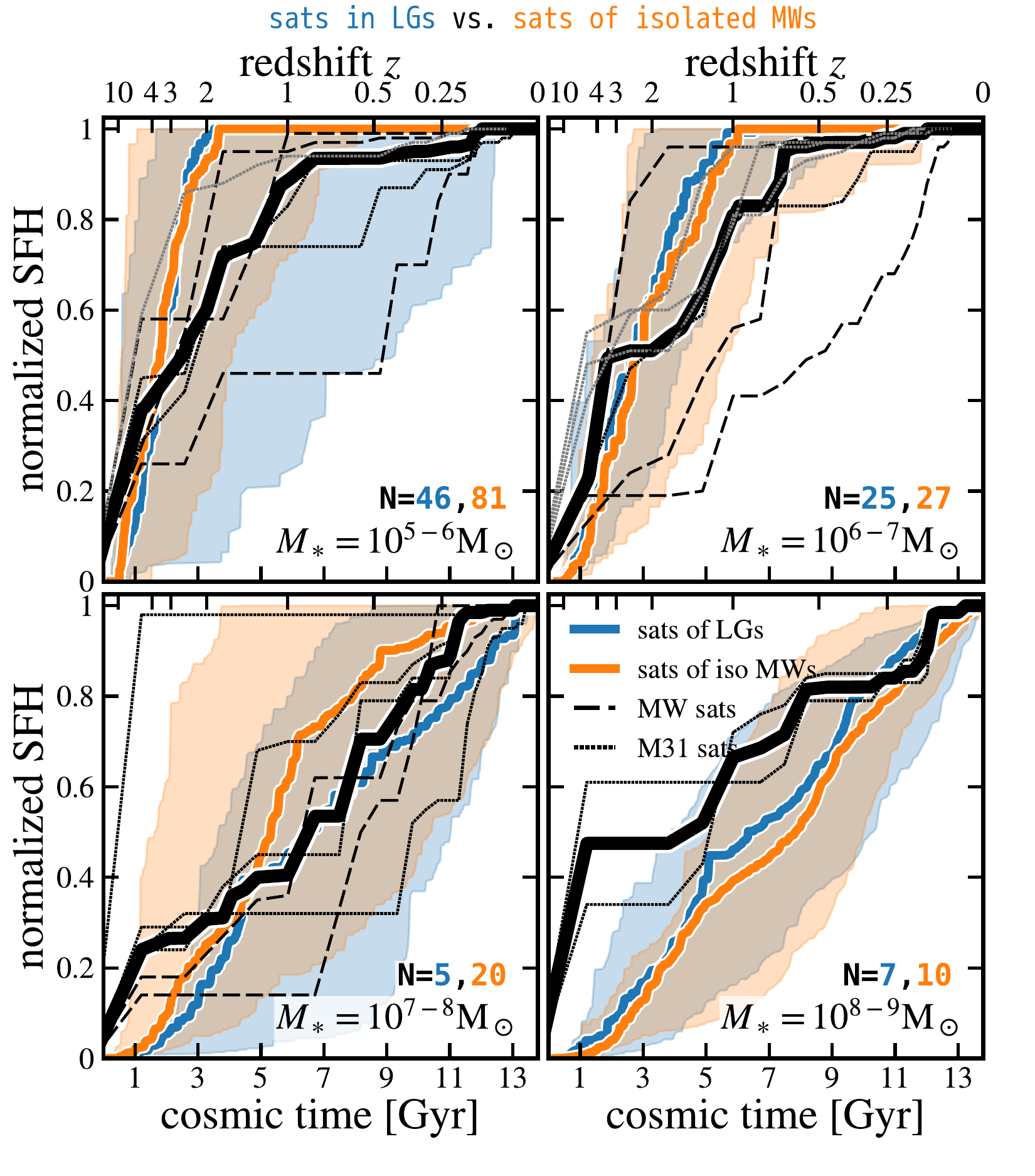}
    \includegraphics[width=\columnwidth]{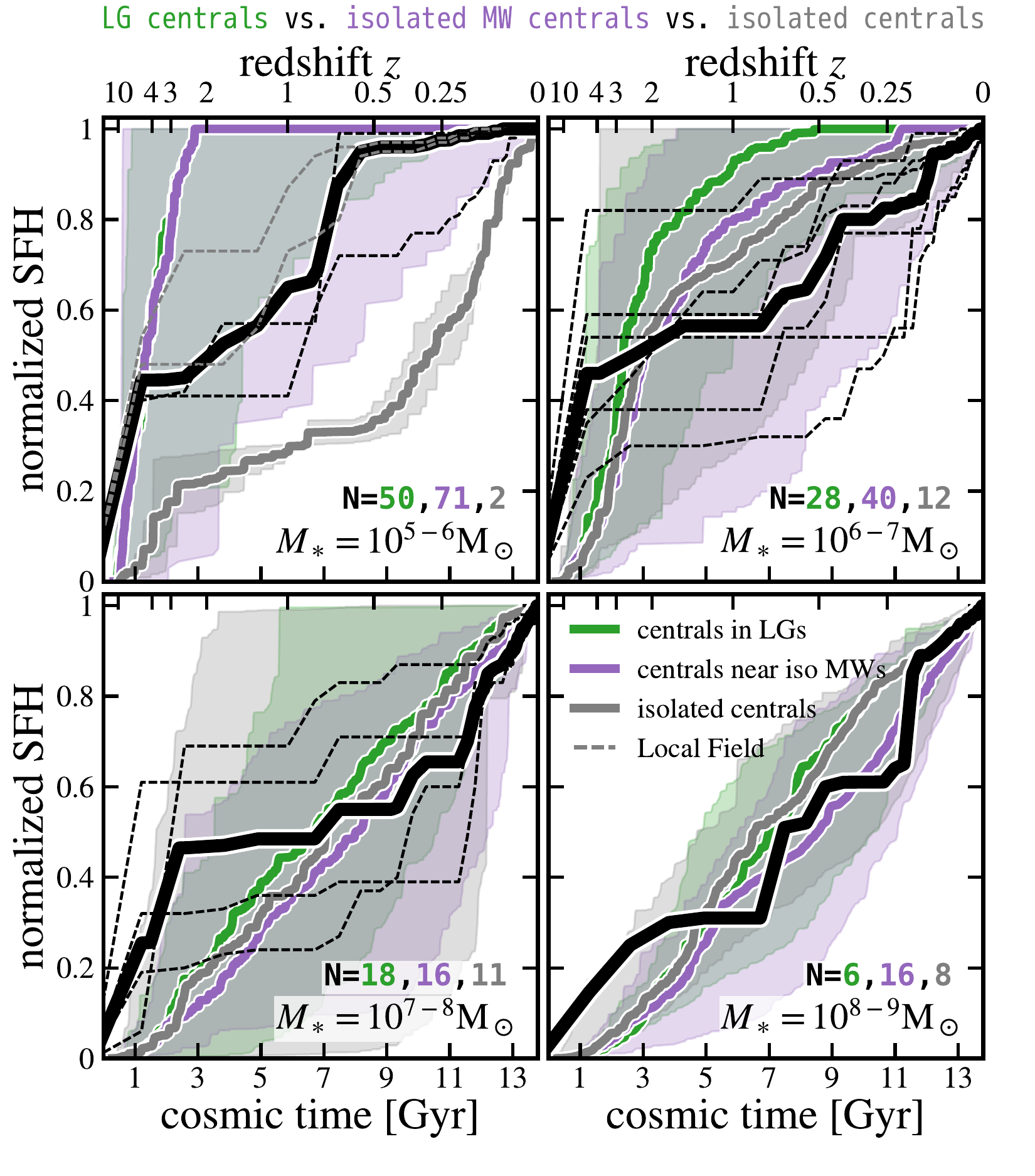}
    \caption{Comparing the SFHs of the simulated dwarf galaxies to 
    observations of the dwarf satellites (left, with line-styles according 
    to the legend) and dwarf centrals (right) in the Local Group.  Observed
    SFHs are taken from \citet[][black lines]{WeiszSFH}
    and \citet[][grey lines]{Skillman2017}.\textsuperscript{\ref{footnote:which_obs}}
    Thick black lines plot the median of the observed galaxies in each bin.
    The simulated medians are identical to Figure~\ref{fig:LGvIso},
    but the shaded regions here indicate the 95\% contours. The 
    simulations reproduce the trends with mass well, and the observations 
    generally lie within the scatter from the simulations, but several of 
    the lower mass galaxies in the Local Group have more late time star 
    formation than those in the FIRE simulations (more akin to the SFHs 
    of our highly isolated dwarf centrals, though not as late-forming).  
    However, we caution that these galaxies are strongly influenced by, 
    e.g., the redshift of reionization, which is relatively early in these 
    simulations ($\zreion\simeq10$; see Appendix~\ref{sec:reion_effects}).}
    \label{fig:LGsVObs}
\end{figure*}

One possibility is that the LG regions are biased and systematically 
collapse earlier, compared to isolated MW or true-field regions. This is 
suggested by Figure~\ref{fig:hostSFHs}, which shows the LG hosts 
preferentially formed their stars $\sim 1-2~\gyr$ earlier than the 
isolated MW hosts. Similarly, comparing the halo growth histories shows 
the paired LG hosts reach half their $z=0$ mass $\sim1~\gyr$ before the 
isolated MW-mass hosts. It is possible this is an artifact of small 
number statistics:  while the number of individual LG dwarfs is large, 
the number of parent LG volumes is only two. In \citet{m12morph},
we consider these and two additional LG volumes (4 total), albeit at lower 
resolution, and find the same average offset in host halo and star formation 
time in the two additional volumes as well. The CLUES project \citep{Gottloeber2010} 
also found a similar effect in dark matter-only simulations of 3 LG analogues. 
However, in dark matter-only simulations of 12 LG volumes, \citet{ELVIS} found 
no significant difference in median LG halo formation times compared to 
isolated MW-mass systems.  Moreover, we find no clear connection between
the half-mass time of a MW-mass host and the quenching times of the dwarf
centrals that evolve nearby:  dwarf centrals around the early-forming 
isolated MWs are not biased relative to those around the later-forming
isolated MWs, nor are the centrals near the late-forming LG-like host
biased relative to those near the other three paired hosts.  Clearly, a 
larger sample of LG volumes with baryonic physics and resolved dwarf
populations is needed to robustly separate baryonic from dark matter 
effects, and (if the latter dominates) whether it is generic to all 
or most LG-type environments.

It is also possible that the offset in the SFHs is either driven by or 
exacerbated by purely baryonic processes.  For example, a combination of 
ram pressure and turbulent viscous stripping can remove cold gas from dwarf 
satellites and shut down star formation, particularly if the hot gas in the 
host halo contains high density clumps \citep{Fillingham2016} or if feedback
within the satellite partially unbinds the cold gas \citep[e.g.][]{Hafen2018}. 
These processes are, in principle, independent of any gravitational 
interactions. The $\sim1-2~\mpc$ density fluctuation required to create an 
LG-like environment may cause the hot halos of the hosts to merge and/or and 
extend further from the central galaxies.  We do find evidence that the 
warm/hot gas ($T\geq10^5$ K) extends further from the hosts in LG-like 
environments than the isolated MW-mass hosts.  Even discounting gas that is 
nearer to the second host in the LG, three of the four LG-like hosts have warm 
gas at $\gtrsim1~\mpc$ while only two of the eight isolated MW-mass hosts 
display similar behavior.  While this intergalactic gas may not be dense enough 
or hot enough to actively strip dwarf galaxies, the comparatively high 
temperatures may inhibit infall of fresh gas (and therefore star formation). 
The relatively early SFHs of the hosts (Figure~\ref{fig:hostSFHs}) could 
also correlate with establishing hot halos at earlier times or with 
supernovae-driven outflows extending further from the host at a 
given time~--~both effects would lead to more efficient quenching 
of nearby dwarf centrals.

\subsection{Comparisons with observations}
Figure~\ref{fig:LGsVObs} repeats Figure~\ref{fig:LGvIso} by again 
plotting the SFHs of our simulated dwarf galaxies, but includes 
95\% contours (rather than 68\%) and adds the observed SFHs of 
MW/M31 satellites and of dwarf centrals in the Local Group.  
Observations are taken from \citet{WeiszSFH} and \citet{Skillman2017}.\footnote{%
    Where the two samples overlap  (And I, II, and III), we adopt the 
    \citet{Skillman2017} SFHs, though the qualitative shapes of the 
    \citet{WeiszSFH} SFHs are identical.
    \label{footnote:which_obs}}
The simulations generally reproduce the trends with mass well:  for 
both satellites and centrals, the observed SFHs are typically within 
the 95\% scatter of the simulations.  We also recover the observed trend
\citep[e.g.][]{Wheeler2014,Wetzel2015b,Weisz2015} wherein massive dwarf
galaxies ($\mstar\simeq10^8$--$10^9\msun$) are nearly impervious to 
quenching: we find with little differences between the SFHs of satellites
and centrals in this mass range.  However, the low mass observed 
satellites have slightly more late time star formation than the 
simulated counterparts, particularly for $\mstar<10^6\msun$.  As 
we discuss below, this discrepancy may be at least partially due 
to resolution, and we note that the two higher resolution, highly 
isolated dwarf centrals in this mass bin have significant star 
formation at much later times~--~so much so, in fact, that they 
under-produce the relative amount of early time star formation 
compared to the observed dwarf centrals in the LG.

\subsection{Caveats}
\label{ssec:caveats}

There are two main caveats to our results:  the resolution of 
the simulations and the time at which the UV background (the 
December 2011 update of the \citealp{FaucherGiguere2009} model) 
reionizes the simulated volume.   We discuss each of these here, 
though we defer a longer discussion of the former to 
Appendix~\ref{sec:resolution}.  Overall, we argue that neither of 
these caveats should impact our main results,  particularly for 
$\mstar\gtrsim 10^6 \msun$ (where the preceding figures 
suggest environmental differences exist).

\subsubsection{Resolution}
Perhaps the most obvious confounding variable in our analysis 
is the diversity of resolutions in the simulations:  though all 
of our runs use identical physics, our isolated dwarf centrals 
vary from $m_i = 30$~--~$4000\msun$, the dwarfs in LG-like environments 
have $m_i = 3500$ or $4000\msun$, and those around isolated 
MW-mass galaxies have $m_i = 4200\msun$ (\run{m12z}) or $7100\msun$ 
(\run{m12b}~--~\run{m12w}).  The dependence of dwarf SFHs on resolution 
in the FIRE-2 simulations is discussed in \citet[][specifically in
\S~4.1.3 and Figure~8]{FIRE2}, but we review those results and 
discuss their impact on our conclusions here.  

In brief, as $m_i$ increases (i.e. resolution decreases), the 
individual bursts of star formation in a given dwarf become 
larger and more violent, as the smallest unit of stars that 
forms is tied to the gas particle mass.  At the lowest 
resolutions, these artificial (i.e. numerical) bursts can become 
large enough that a single burst removes all of the gas from
a dwarf, which can permanently shut off star formation in that
dwarf and tends to reduce the amount of late time star formation 
on average.  Nonetheless, \emph{total} stellar masses remain 
remarkably consistent -- typically to within $\sim20\%$, and 
within a factor of $\sim3$ even with only two star particles 
in the galaxy.  \citet{FIRE2} and Appendix~\ref{sec:resolution} 
demonstrate that the SFH of a $\sim10^6\msun$ galaxy is reasonably 
well resolved at $m_i \simeq 2100\msun$, though the $m_i = 30\msun$ 
simulation we use here is shifted to slightly later times.  
Therefore, the SFHs of the lower mass ($\mstar\lesssim10^7\msun$) 
dwarf galaxies taken from MW/LG environments, as well as those 
in the \citet{Graus2019} sample, may be somewhat under-resolved. 

Correcting for these artificially concentrated bursts should 
actually enhance the differences we find between LG-like 
environments and those of isolated MW-mass hosts, however.  
Dwarf galaxies in the LG-like simulations tend to form 
\emph{earlier} than their analogues around isolated MW-mass 
hosts, even though the LG simulations are at slightly 
\emph{higher} resolutions.  Therefore, were the samples 
to be run at identical resolutions, we would expect to find 
an even larger difference between isolated MWs and LG-like 
environments, with the former exhibiting even more extended SFHs.  
We cannot, though, rule out the possibility that the offset 
between the highly isolated dwarf centrals and dwarf galaxies 
around MW-mass host(s), wherein the latter form earlier, is 
exaggerated by resolution.  However, we show in
Appendix~\ref{sec:resolution} that the $m_i = 4000\msun$
\citet{Graus2019} galaxies form later than those in the 
$m_i=500\msun$ \citet{Fitts2017} sample, suggesting that 
resolution effects are sub-dominant to mass trends and 
galaxy-to-galaxy scatter.  We also note that correcting 
for these resolution trends would tend to shift the 
simulated SFHs more in line with the observations.  

\subsubsection{Time of reionization}
By default (and in all the runs analyzed here), the FIRE-2 
simulations adopt the \citet{FaucherGiguere2009} photo-ionizing 
background to capture the meta-galactic UV photons responsible 
for cosmic reionization.  As mentioned above, that model was 
designed to match the WMAP-7 optical depth, corresponding to a 
reionization redshift of $z\simeq10$ \citep{Komatsu2011}, in 
contrast with the most recent constraints on the midpoint of 
reionization from the Planck mission of $z=7.68\pm0.79$ 
\citep{Planck2018}.  We emphasize, though, that all of the dwarf 
galaxies analyzed here were simulated with an identical 
photo-ionizing background.  Therefore, we expect that a later 
reionizing background would shift all of our SFHs (at a given 
mass) in the same manner; that is, the relative comparisons 
between the different environmental samples should be robust 
to the time of reionization.  

However, this discussion ignores the effects of ``patchy'' 
reionization.  Both simulations \citep[e.g.][]{Trac2007,Ocvirk2016,
Chen2017} and observations \citep[e.g.][]{Pentericci2014,Zheng2017} 
suggest some sections of the Universe may take until $z\sim6.5$
to complete reionization.  If reionization is highly patchy, and 
if the proto-MW galaxies contribute significantly to the local 
reionizing field, then we would naively expect that LG-like 
environments should reionize before those around isolated MWs, 
which should reionize before regions that host highly isolated 
dwarf galaxies \citep[e.g.][]{Alvarez2009,Lunnan2012}.  Therefore, 
at the masses where reionization interferes with star formation 
(roughly $\mhalo(z=\zreion)\lesssim10^9\msun$; 
\citealp{Dawoodbhoy2018}), we would expect more early-time 
star formation in the isolated dwarf galaxies and less in those 
evolving in LG-like environments.  As we discuss in Appendix~\ref{sec:reion_effects} 
(and show explicitly in Figure~\ref{fig:m12i-sats-reion}), a 
later $\zreion$ leads to a smaller fraction of late-time star 
formation, i.e. the normalized SFHs are shifted to earlier times.  
Consequently, patchy reionization could act to smear out the 
differences that we find between the three environments.  However, 
reionization has the strongest impact on galaxies with 
$\mstar\lesssim10^6\msun$ \citep[e.g.][]{Wheeler2015,Wheeler2018}, 
and we find evidence of environmental variations at 
$\mstar=10^6$~--~$10^7\msun$.  That is, the strongest differences 
exist for galaxies that should not be strongly impacted by the 
timing and patchiness of reionization.

Finally, in the process of preparing this manuscript, we discovered 
that an external heating term designed to mimic cosmic rays in the 
interstellar medium of MW-mass galaxies was being improperly applied 
to the inter\emph{galactic} medium at extremely high redshift 
($z\gtrsim20$; also see \citealp{Su2018}). Like the too-early 
reionization, this would act to suppress star formation at early 
times, such that a greater fraction of star formation instead occurs
at later times.  Correcting this mistake should therefore shift our 
SFHs to earlier times overall, since they represent the fractional 
mass formed by a given time.  Internal testing indicates that this 
extraneous heating term has less than a $10\%$ impact on the shapes 
of SFHs for $\mstar\sim10^6\msun$, with the strength of the effect 
scaling inversely with galaxy mass.  Therefore, we expect this error 
to have a marginal impact for our three higher mass bins, and we 
again emphasize that, even at lower masses, \emph{all} of our runs 
include this spurious heating term~--~correcting it should impact 
all of our SFHs in the same manner, and therefore only impact 
our conclusions with respect to the observations.

\section{Conclusions}
\label{sec:conclusions}

We have used a set of $\simeq500$ dwarf galaxies~--~taken from FIRE-2
simulations of LG-like pairs of Milky Way-mass hosts separated by 
$\lesssim1~\mpc$, from isolated (single) MW-mass galaxies that are 
at least $\gtrsim3~\mpc$ from any other MW-mass systems, and from
low density regions that contain no MW-mass hosts in the simulation 
volume~--~to explore how the shapes of dwarf star formation histories 
vary with environment (both in terms of the number of MW-mass hosts 
nearby and the $z=0$ distance to the nearest such galaxy).  Our main 
conclusions are:

\begin{itemize}
\item Even at fixed mass and environment, there is a large 
degree of scatter in star formation histories, with the full 
sample often spanning $\mstar(z)/\mstar(z=0) \simeq 0.2$~--~$1$
at fixed cosmic time. 

\item Nonetheless, the trends with mass in the median star 
formation histories are robust:  the fraction of stars formed 
at late times (roughly defined as cosmic time $t\geq5~\gyr$) 
increases with the $z=0$ stellar mass of the dwarf, nearly 
independent of environment (Figures~\ref{fig:allSFHs}~and~\ref{fig:moments}).

\item The satellites (defined as galaxies within $300~\kpc$ of
a MW-mass host) of isolated MW-mass galaxies tend to form
their stars earlier than equivalent dwarf centrals (non-satellite 
galaxies that are near a MW-mass host, but more than $300~\kpc$ 
away from that host today), consistent with a picture where 
interactions with MW-mass hosts inhibit star formation 
(Figure~\ref{fig:satVfield}).

\item Satellites in LG-like pairs have nearly identical SFHs 
to those of satellites around isolated MW-mass galaxies 
(Figure~\ref{fig:LGvIso}).

\item Dwarf central galaxies that evolve in LG-like environments 
have SFHs that are relatively similar to their satellite counterparts; 
that is, they contain older stars than central galaxies of 
similar masses around isolated MW-mass galaxies 
(Figures~\ref{fig:satVfield}~and~\ref{fig:LGvIso}).

\item Highly isolated dwarf galaxies with $\mstar\leq10^7\msun$ 
form even later than the centrals around isolated MW-mass galaxies, 
suggesting a trend with the number of nearby MWs 
(Figures~\ref{fig:LGvIso}~and~\ref{fig:moments}).  However, the 
difference at $\mstar\leq10^6\msun$ may be exaggerated by resolution, 
and the offset at $\mstar=10^{6-7}\msun$ relative to the centrals 
around isolated MW-mass hosts is not statistically significant, 
though the offset compared to the LG dwarf centrals is.

\item A statistical analysis of several summary statistics
of each SFH (specifically, the time when each galaxy reaches 
10\%, 50\%, and 90\% of its mass at $z=0$), as a function of 
environment, indicates that~--~for $\mstar=10^{6-7}\msun$~--~the
satellites of isolated MWs and the dwarf galaxies in LGs
(satellites or centrals) do not significantly differ from 
one another, though highly isolated dwarf centrals and the
dwarf centrals around isolated (non-paired) MWs do differ
strongly from the former three environments.  The comparison
between LG centrals and centrals around isolated MWs is 
inconclusive, but the AD test suggests that the former are 
more similar to the LG satellites than to the isolated-MW 
centrals (Figure~\ref{fig:ADmatrix}).

\item The simulations broadly reproduce the observed SFHs of both 
satellite and central dwarf galaxies in the LG.  While they 
underproduce the amount of late time star formation in our lowest 
mass dwarfs ($\mstar=10^{5-6}\msun$), the disagreement is at least 
qualitatively consistent with resolution artifacts.  However,
there is a slight tension in that the observed SFHs of the LG
centrals are slightly more consistent with the later forming
centrals in simulations of isolated MWs, rather than with the
(relatively) early forming centrals in simulations targeting 
LGs (Figure~\ref{fig:LGsVObs}).

\end{itemize}

Our results suggest that the MW-mass galaxies in the FIRE-2 
simulations affect star formation in the dwarfs around them
(though potentially indirectly), even when those dwarfs are 
not satellites at $z=0$.  Therefore, caution should be taken 
when comparing simulations of dwarf galaxies that do not 
include any MW-mass galaxies, particularly if the properties 
under consideration are sensitive to the timing of star formation.
Further work is required to solidify the significance of the 
differences (via increased sample sizes) and to identify their
direct causes.

\section*{Acknowledgments}

The authors thank Cameron Hummels, Peter Behroozi, Stephanie Tonnesen,
and Zach Hafen for valuable discussions, Dan Weisz and Evan Skillman 
for making publicly available the observed star formation histories, 
and Oliver Hahn and Peter Behroozi for making \texttt{MUSIC} and 
\texttt{Rockstar} public, respectively.  This work was performed in 
part at the Aspen Center for Physics, which is supported by National 
Science Foundation grant PHY-1607611.

Support for SGK was provided by NASA through Einstein Postdoctoral Fellowship grant number PF5-160136 awarded by the Chandra X-ray Center, which is operated by the Smithsonian Astrophysical Observatory for NASA under contract NAS8-03060.
SGK, AW, JSB, KEB, RES, MBK, and CAFG were supported by NASA through 
ATP \#80NSSC18K0562, 80NSSC18K1097,
HST-GO-12914, HST-GO-14734, HST-GO-14191, 
HST-AR-13888, HST-AR-13896, HST-AR-13921, HST-AR-14282, HST-AR-14554, HST-AR-15006, HST-AR-15057, 
JPL 1589742, 
NNX17AG29G, NNX15AB22G, and 17-ATP17-0067, 
many of which are awarded by STScI, which is operated by the Association of Universities for Research in Astronomy (AURA), Inc., under NASA contract NAS5-26555.
SGK, PFH, JSB, AG, KEB, MBK, AF, TKC, DK, and CAFG were supported by the NSF through
CAREER grant \#1455342, CAREER grant AST-1752913, CAREER grant AST-1652522, Collaborative Research Grant \#1715847, 
AST-1517226, AST-1517491, AST-1518291, AST-1715101, AST-1715216, and an NSF Graduate Research Fellowship.  
Additional support was provided by the Moore Center for Theoretical Cosmology and Physics at Caltech, the Alfred P. Sloan Research Fellowship, a Berkeley graduate fellowship, a Hellman award for graduate study, the Lee A. DuBridge Postdoctoral Scholarship in Astrophysics, the Swiss National Science Foundation (grant \#157591), and the Cottrell Scholarship Award from the Research Corporation for Science Advancement.  The Flatiron Institute is supported by the Simons Foundation.  

Numerical calculations were run on the Caltech compute cluster ``Wheeler,''
allocations from XSEDE TG-AST130039 and PRAC NSF.1713353 supported by the
NSF, NASA HEC SMD-16-7223 and SMD-16-7592, and High Performance Computing
at Los Alamos National Labs.  This work also made use of \texttt{Astropy},
a community-developed core Python package for Astronomy \citep{Astropy,Astropy2},
\texttt{matplotlib} \citep{Matplotlib}, \texttt{numpy} \citep{numpy},
\texttt{scipy} \citep{scipy}, \texttt{ipython} \citep{ipython}, \texttt{yt}
\citep{yt}, \texttt{ytree} \citep{ytree}, and NASA's Astrophysics Data System.  

\bibliographystyle{mnras}
\bibliography{fire_sfhs}

\begin{thebibliography}{}
\makeatletter
\relax
\def\mn@urlcharsother{\let\do\@makeother \do\$\do\&\do\#\do\^\do\_\do\%\do\~}
\def\mn@doi{\begingroup\mn@urlcharsother \@ifnextchar [ {\mn@doi@}
  {\mn@doi@[]}}
\def\mn@doi@[#1]#2{\def\@tempa{#1}\ifx\@tempa\@empty \href
  {http://dx.doi.org/#2} {doi:#2}\else \href {http://dx.doi.org/#2} {#1}\fi
  \endgroup}
\def\mn@eprint#1#2{\mn@eprint@#1:#2::\@nil}
\def\mn@eprint@arXiv#1{\href {http://arxiv.org/abs/#1} {{\tt arXiv:#1}}}
\def\mn@eprint@dblp#1{\href {http://dblp.uni-trier.de/rec/bibtex/#1.xml}
  {dblp:#1}}
\def\mn@eprint@#1:#2:#3:#4\@nil{\def\@tempa {#1}\def\@tempb {#2}\def\@tempc
  {#3}\ifx \@tempc \@empty \let \@tempc \@tempb \let \@tempb \@tempa \fi \ifx
  \@tempb \@empty \def\@tempb {arXiv}\fi \@ifundefined
  {mn@eprint@\@tempb}{\@tempb:\@tempc}{\expandafter \expandafter \csname
  mn@eprint@\@tempb\endcsname \expandafter{\@tempc}}}

\bibitem[\protect\citeauthoryear{{Alvarez}, {Busha}, {Abel}  \&
  {Wechsler}}{{Alvarez} et~al.}{2009}]{Alvarez2009}
{Alvarez} M.~A.,  {Busha} M.,  {Abel} T.,   {Wechsler} R.~H.,  2009, \mn@doi
  [\apjl] {10.1088/0004-637X/703/2/L167}, \href
  {http://adsabs.harvard.edu/abs/2009ApJ...703L.167A} {703, L167}

\bibitem[\protect\citeauthoryear{Anderson \& Darling}{Anderson \&
  Darling}{1954}]{Anderson1956:ADTest}
Anderson T.~W.,  Darling D.~A.,  1954, Journal of the American Statistical
  Association, 49, 765

\bibitem[\protect\citeauthoryear{{Angl{\'e}s-Alc{\'a}zar},
  {Faucher-Gigu{\`e}re}, {Kere{\v s}}, {Hopkins}, {Quataert}  \&
  {Murray}}{{Angl{\'e}s-Alc{\'a}zar} et~al.}{2017}]{AnglesAcazar2017}
{Angl{\'e}s-Alc{\'a}zar} D.,  {Faucher-Gigu{\`e}re} C.-A.,  {Kere{\v s}} D.,
  {Hopkins} P.~F.,  {Quataert} E.,   {Murray} N.,  2017, \mn@doi [\mnras]
  {10.1093/mnras/stx1517}, \href
  {http://adsabs.harvard.edu/abs/2017MNRAS.470.4698A} {470, 4698}

\bibitem[\protect\citeauthoryear{{Astropy Collaboration} et~al.,}{{Astropy
  Collaboration} et~al.}{2013}]{Astropy}
{Astropy Collaboration} et~al., 2013, \mn@doi [\aap]
  {10.1051/0004-6361/201322068}, \href
  {http://adsabs.harvard.edu/abs/2013A%26A...558A..33A} {558, A33}

\bibitem[\protect\citeauthoryear{{Behroozi}, {Wechsler}  \& {Wu}}{{Behroozi}
  et~al.}{2013}]{rockstar}
{Behroozi} P.~S.,  {Wechsler} R.~H.,   {Wu} H.-Y.,  2013, \mn@doi [\apj]
  {10.1088/0004-637X/762/2/109}, \href
  {http://adsabs.harvard.edu/abs/2013ApJ...762..109B} {762, 109}

\bibitem[\protect\citeauthoryear{{Ben{\'{\i}}tez-Llambay}, {Navarro}, {Abadi},
  {Gottl{\"o}ber}, {Yepes}, {Hoffman}  \& {Steinmetz}}{{Ben{\'{\i}}tez-Llambay}
  et~al.}{2015}]{Benitez-Llambay2015}
{Ben{\'{\i}}tez-Llambay} A.,  {Navarro} J.~F.,  {Abadi} M.~G.,  {Gottl{\"o}ber}
  S.,  {Yepes} G.,  {Hoffman} Y.,   {Steinmetz} M.,  2015, \mn@doi [\mnras]
  {10.1093/mnras/stv925}, \href
  {http://adsabs.harvard.edu/abs/2015MNRAS.450.4207B} {450, 4207}

\bibitem[\protect\citeauthoryear{{Boylan-Kolchin}, {Weisz}, {Johnson},
  {Bullock}, {Conroy}  \& {Fitts}}{{Boylan-Kolchin}
  et~al.}{2015}]{Boylan-Kolchin2015}
{Boylan-Kolchin} M.,  {Weisz} D.~R.,  {Johnson} B.~D.,  {Bullock} J.~S.,
  {Conroy} C.,   {Fitts} A.,  2015, \mn@doi [\mnras] {10.1093/mnras/stv1736},
  \href {http://adsabs.harvard.edu/abs/2015MNRAS.453.1503B} {453, 1503}

\bibitem[\protect\citeauthoryear{{Bozek} et~al.,}{{Bozek}
  et~al.}{2019}]{Bozek2018}
{Bozek} B.,  et~al., 2019, \mn@doi [\mnras] {10.1093/mnras/sty3300}, \href
  {https://ui.adsabs.harvard.edu/\#abs/2019MNRAS.483.4086B} {483, 4086}

\bibitem[\protect\citeauthoryear{{Brooks} \& {Zolotov}}{{Brooks} \&
  {Zolotov}}{2014}]{BrooksZolotov2012}
{Brooks} A.~M.,  {Zolotov} A.,  2014, \mn@doi [\apj]
  {10.1088/0004-637X/786/2/87}, \href
  {http://adsabs.harvard.edu/abs/2014ApJ...786...87B} {786, 87}

\bibitem[\protect\citeauthoryear{{Brown} et~al.,}{{Brown}
  et~al.}{2014}]{Brown2014}
{Brown} T.~M.,  et~al., 2014, \memsai, \href
  {http://adsabs.harvard.edu/abs/2014MmSAI..85..493B} {85, 493}

\bibitem[\protect\citeauthoryear{{Cole} et~al.,}{{Cole}
  et~al.}{2007}]{Cole2007}
{Cole} A.~A.,  et~al., 2007, \mn@doi [\apj] {10.1086/516711}, \href
  {https://ui.adsabs.harvard.edu/#abs/2007ApJ...659L..17C} {659, L17}

\bibitem[\protect\citeauthoryear{{Cole}, {Weisz}, {Dolphin}, {Skillman},
  {McConnachie}, {Brooks}  \& {Leaman}}{{Cole} et~al.}{2014}]{Cole2014}
{Cole} A.~A.,  {Weisz} D.~R.,  {Dolphin} A.~E.,  {Skillman} E.~D.,
  {McConnachie} A.~W.,  {Brooks} A.~M.,   {Leaman} R.,  2014, \mn@doi [\apj]
  {10.1088/0004-637X/795/1/54}, \href
  {https://ui.adsabs.harvard.edu/#abs/2014ApJ...795...54C} {795, 54}

\bibitem[\protect\citeauthoryear{{Col{\'{\i}}n}, {Avila-Reese},
  {Gonz{\'a}lez-Samaniego}  \& {Vel{\'a}zquez}}{{Col{\'{\i}}n}
  et~al.}{2015}]{Colin2015}
{Col{\'{\i}}n} P.,  {Avila-Reese} V.,  {Gonz{\'a}lez-Samaniego} A.,
  {Vel{\'a}zquez} H.,  2015, \mn@doi [\apj] {10.1088/0004-637X/803/1/28}, \href
  {http://adsabs.harvard.edu/abs/2015ApJ...803...28C} {803, 28}

\bibitem[\protect\citeauthoryear{{Dawoodbhoy} et~al.,}{{Dawoodbhoy}
  et~al.}{2018}]{Dawoodbhoy2018}
{Dawoodbhoy} T.,  et~al., 2018, \mn@doi [\mnras] {10.1093/mnras/sty1945}, \href
  {http://adsabs.harvard.edu/abs/2018MNRAS.480.1740D} {480, 1740}

\bibitem[\protect\citeauthoryear{{Deason}, {Wetzel}  \&
  {Garrison-Kimmel}}{{Deason} et~al.}{2014}]{Deason2014}
{Deason} A.,  {Wetzel} A.,   {Garrison-Kimmel} S.,  2014, \mn@doi [\apj]
  {10.1088/0004-637X/794/2/115}, \href
  {http://adsabs.harvard.edu/abs/2014ApJ...794..115D} {794, 115}

\bibitem[\protect\citeauthoryear{{Digby} et~al.,}{{Digby}
  et~al.}{2018}]{Digby2018}
{Digby} R.,  et~al., 2018, preprint, \href
  {http://adsabs.harvard.edu/abs/2018arXiv181205669D} {} (\mn@eprint {arXiv}
  {1812.05669})

\bibitem[\protect\citeauthoryear{Dunn}{Dunn}{1961}]{Dunn1961}
Dunn O.~J.,  1961, \mn@doi [Journal of the American Statistical Association]
  {10.1080/01621459.1961.10482090}, 56, 52

\bibitem[\protect\citeauthoryear{{Einasto}, {Saar}, {Kaasik}  \&
  {Chernin}}{{Einasto} et~al.}{1974}]{Einasto1974}
{Einasto} J.,  {Saar} E.,  {Kaasik} A.,   {Chernin} A.~D.,  1974, \mn@doi
  [\nat] {10.1038/252111a0}, \href
  {http://adsabs.harvard.edu/abs/1974Natur.252..111E} {252, 111}

\bibitem[\protect\citeauthoryear{{El-Badry} et~al.,}{{El-Badry}
  et~al.}{2018a}]{ElBadry2018a}
{El-Badry} K.,  et~al., 2018a, \mn@doi [\mnras] {10.1093/mnras/stx2482}, \href
  {https://ui.adsabs.harvard.edu/#abs/2018MNRAS.473.1930E} {473, 1930}

\bibitem[\protect\citeauthoryear{{El-Badry} et~al.,}{{El-Badry}
  et~al.}{2018b}]{ElBadry2018}
{El-Badry} K.,  et~al., 2018b, \mn@doi [\mnras] {10.1093/mnras/sty730}, \href
  {http://adsabs.harvard.edu/abs/2018MNRAS.477.1536E} {477, 1536}

\bibitem[\protect\citeauthoryear{{Emerick}, {Mac Low}, {Grcevich}  \&
  {Gatto}}{{Emerick} et~al.}{2016}]{Emerick2016}
{Emerick} A.,  {Mac Low} M.-M.,  {Grcevich} J.,   {Gatto} A.,  2016, \mn@doi
  [\apj] {10.3847/0004-637X/826/2/148}, \href
  {http://adsabs.harvard.edu/abs/2016ApJ...826..148E} {826, 148}

\bibitem[\protect\citeauthoryear{{Escala} et~al.,}{{Escala}
  et~al.}{2018}]{Escala2017}
{Escala} I.,  et~al., 2018, \mn@doi [\mnras] {10.1093/mnras/stx2858}, \href
  {http://adsabs.harvard.edu/abs/2018MNRAS.474.2194E} {474, 2194}

\bibitem[\protect\citeauthoryear{{Fattahi} et~al.,}{{Fattahi}
  et~al.}{2016}]{Fattahi2016}
{Fattahi} A.,  et~al., 2016, \mn@doi [\mnras] {10.1093/mnras/stv2970}, \href
  {http://adsabs.harvard.edu/abs/2016MNRAS.457..844F} {457, 844}

\bibitem[\protect\citeauthoryear{{Faucher-Gigu{\`e}re}, {Lidz}, {Zaldarriaga}
  \& {Hernquist}}{{Faucher-Gigu{\`e}re} et~al.}{2009}]{FaucherGiguere2009}
{Faucher-Gigu{\`e}re} C.-A.,  {Lidz} A.,  {Zaldarriaga} M.,   {Hernquist} L.,
  2009, \mn@doi [\apj] {10.1088/0004-637X/703/2/1416}, \href
  {http://adsabs.harvard.edu/abs/2009ApJ...703.1416F} {703, 1416}

\bibitem[\protect\citeauthoryear{{Fillingham}, {Cooper}, {Wheeler},
  {Garrison-Kimmel}, {Boylan-Kolchin}  \& {Bullock}}{{Fillingham}
  et~al.}{2015}]{Fillingham2015}
{Fillingham} S.~P.,  {Cooper} M.~C.,  {Wheeler} C.,  {Garrison-Kimmel} S.,
  {Boylan-Kolchin} M.,   {Bullock} J.~S.,  2015, \mn@doi [\mnras]
  {10.1093/mnras/stv2058}, \href
  {http://adsabs.harvard.edu/abs/2015MNRAS.454.2039F} {454, 2039}

\bibitem[\protect\citeauthoryear{{Fillingham}, {Cooper}, {Pace},
  {Boylan-Kolchin}, {Bullock}, {Garrison-Kimmel}  \& {Wheeler}}{{Fillingham}
  et~al.}{2016}]{Fillingham2016}
{Fillingham} S.~P.,  {Cooper} M.~C.,  {Pace} A.~B.,  {Boylan-Kolchin} M.,
  {Bullock} J.~S.,  {Garrison-Kimmel} S.,   {Wheeler} C.,  2016, \mn@doi
  [\mnras] {10.1093/mnras/stw2131}, \href
  {http://adsabs.harvard.edu/abs/2016MNRAS.463.1916F} {463, 1916}

\bibitem[\protect\citeauthoryear{{Fillingham}, {Cooper}, {Boylan-Kolchin},
  {Bullock}, {Garrison-Kimmel}  \& {Wheeler}}{{Fillingham}
  et~al.}{2018}]{Fillingham2018}
{Fillingham} S.~P.,  {Cooper} M.~C.,  {Boylan-Kolchin} M.,  {Bullock} J.~S.,
  {Garrison-Kimmel} S.,   {Wheeler} C.,  2018, \mn@doi [\mnras]
  {10.1093/mnras/sty958}, \href
  {http://adsabs.harvard.edu/abs/2018MNRAS.477.4491F} {477, 4491}

\bibitem[\protect\citeauthoryear{{Fitts} et~al.,}{{Fitts}
  et~al.}{2017}]{Fitts2017}
{Fitts} A.,  et~al., 2017, \mn@doi [\mnras] {10.1093/mnras/stx1757}, \href
  {http://adsabs.harvard.edu/abs/2017MNRAS.471.3547F} {471, 3547}

\bibitem[\protect\citeauthoryear{{Fitts} et~al.,}{{Fitts}
  et~al.}{2018}]{Fitts2018}
{Fitts} A.,  et~al., 2018, \mn@doi [\mnras] {10.1093/mnras/sty1488}, \href
  {https://ui.adsabs.harvard.edu/#abs/2018MNRAS.479..319F} {479, 319}

\bibitem[\protect\citeauthoryear{{Gallart} et~al.,}{{Gallart}
  et~al.}{2015}]{Gallart2015}
{Gallart} C.,  et~al., 2015, \mn@doi [\apjl] {10.1088/2041-8205/811/2/L18},
  \href {http://adsabs.harvard.edu/abs/2015ApJ...811L..18G} {811, L18}

\bibitem[\protect\citeauthoryear{{Garrison-Kimmel}, {Boylan-Kolchin}, {Bullock}
   \& {Lee}}{{Garrison-Kimmel} et~al.}{2014}]{ELVIS}
{Garrison-Kimmel} S.,  {Boylan-Kolchin} M.,  {Bullock} J.~S.,   {Lee} K.,
  2014, \mn@doi [\mnras] {10.1093/mnras/stt2377}, \href
  {http://adsabs.harvard.edu/abs/2014MNRAS.438.2578G} {438, 2578}

\bibitem[\protect\citeauthoryear{{Garrison-Kimmel} et~al.,}{{Garrison-Kimmel}
  et~al.}{2017}]{GKDisk}
{Garrison-Kimmel} S.,  et~al., 2017, \mn@doi [\mnras] {10.1093/mnras/stx1710},
  \href {http://adsabs.harvard.edu/abs/2017MNRAS.471.1709G} {471, 1709}

\bibitem[\protect\citeauthoryear{{Garrison-Kimmel} et~al.,}{{Garrison-Kimmel}
  et~al.}{2018a}]{GK2018}
{Garrison-Kimmel} S.,  et~al., 2018a, preprint, \href
  {http://adsabs.harvard.edu/abs/2018arXiv180604143G} {} (\mn@eprint {arXiv}
  {1806.04143})

\bibitem[\protect\citeauthoryear{{Garrison-Kimmel} et~al.,}{{Garrison-Kimmel}
  et~al.}{2018b}]{m12morph}
{Garrison-Kimmel} S.,  et~al., 2018b, \mn@doi [\mnras] {10.1093/mnras/sty2513},
  \href {http://adsabs.harvard.edu/abs/2018MNRAS.481.4133G} {481, 4133}

\bibitem[\protect\citeauthoryear{{Geha}, {Blanton}, {Yan}  \& {Tinker}}{{Geha}
  et~al.}{2012}]{Geha2012}
{Geha} M.,  {Blanton} M.~R.,  {Yan} R.,   {Tinker} J.~L.,  2012, \mn@doi [\apj]
  {10.1088/0004-637X/757/1/85}, \href
  {http://adsabs.harvard.edu/abs/2012ApJ...757...85G} {757, 85}

\bibitem[\protect\citeauthoryear{{Geha} et~al.,}{{Geha} et~al.}{2017}]{SAGA1}
{Geha} M.,  et~al., 2017, \mn@doi [\apj] {10.3847/1538-4357/aa8626}, \href
  {http://adsabs.harvard.edu/abs/2017ApJ...847....4G} {847, 4}

\bibitem[\protect\citeauthoryear{{Gottloeber}, {Hoffman}  \&
  {Yepes}}{{Gottloeber} et~al.}{2010}]{Gottloeber2010}
{Gottloeber} S.,  {Hoffman} Y.,   {Yepes} G.,  2010, preprint, \href
  {http://adsabs.harvard.edu/abs/2010arXiv1005.2687G} {} (\mn@eprint {arXiv}
  {1005.2687})

\bibitem[\protect\citeauthoryear{{Governato} et~al.,}{{Governato}
  et~al.}{2015}]{Governato2015}
{Governato} F.,  et~al., 2015, \mn@doi [\mnras] {10.1093/mnras/stu2720}, \href
  {http://adsabs.harvard.edu/abs/2015MNRAS.448..792G} {448, 792}

\bibitem[\protect\citeauthoryear{{Grand} et~al.,}{{Grand}
  et~al.}{2017}]{Grand2017}
{Grand} R.~J.~J.,  et~al., 2017, \mn@doi [\mnras] {10.1093/mnras/stx071}, \href
  {http://adsabs.harvard.edu/abs/2017MNRAS.467..179G} {467, 179}

\bibitem[\protect\citeauthoryear{{Graus} et~al.,}{{Graus}
  et~al.}{2019}]{Graus2019}
{Graus} A.~S.,  et~al., 2019, preprint, \href
  {http://adsabs.harvard.edu/abs/2019arXiv190105487G} {} (\mn@eprint {arXiv}
  {1901.05487})

\bibitem[\protect\citeauthoryear{{Gunn} \& {Gott}}{{Gunn} \&
  {Gott}}{1972}]{Gunn1972}
{Gunn} J.~E.,  {Gott} III J.~R.,  1972, \mn@doi [\apj] {10.1086/151605}, \href
  {http://adsabs.harvard.edu/abs/1972ApJ...176....1G} {176, 1}

\bibitem[\protect\citeauthoryear{Gupta, Mathur, Krongold, Nicastro  \&
  Galeazzi}{Gupta et~al.}{2012}]{Gupta2012}
Gupta A.,  Mathur S.,  Krongold Y.,  Nicastro F.,   Galeazzi M.,  2012, The
  Astrophysical Journal Letters, 756, L8

\bibitem[\protect\citeauthoryear{{Hafen} et~al.,}{{Hafen}
  et~al.}{2018}]{Hafen2018}
{Hafen} Z.,  et~al., 2018, arXiv e-prints, \href
  {http://adsabs.harvard.edu/abs/2018arXiv181111753H} {}

\bibitem[\protect\citeauthoryear{{Hahn} \& {Abel}}{{Hahn} \&
  {Abel}}{2011}]{MUSIC}
{Hahn} O.,  {Abel} T.,  2011, \mn@doi [\mnras]
  {10.1111/j.1365-2966.2011.18820.x}, \href
  {http://adsabs.harvard.edu/abs/2011MNRAS.415.2101H} {415, 2101}

\bibitem[\protect\citeauthoryear{{Hester}}{{Hester}}{2006}]{Hester2006}
{Hester} J.~A.,  2006, \mn@doi [\apj] {10.1086/505614}, \href
  {http://adsabs.harvard.edu/abs/2006ApJ...647..910H} {647, 910}

\bibitem[\protect\citeauthoryear{{Hopkins}}{{Hopkins}}{2015}]{GIZMO}
{Hopkins} P.~F.,  2015, \mn@doi [\mnras] {10.1093/mnras/stv195}, \href
  {http://adsabs.harvard.edu/abs/2015MNRAS.450...53H} {450, 53}

\bibitem[\protect\citeauthoryear{{Hopkins}}{{Hopkins}}{2017}]{Hopkinsmetaldiff}
{Hopkins} P.~F.,  2017, \mn@doi [\mnras] {10.1093/mnras/stw3306}, \href
  {http://adsabs.harvard.edu/abs/2017MNRAS.466.3387H} {466, 3387}

\bibitem[\protect\citeauthoryear{{Hopkins}, {Narayanan}  \& {Murray}}{{Hopkins}
  et~al.}{2013}]{Hopkins2013sf_criteria}
{Hopkins} P.~F.,  {Narayanan} D.,   {Murray} N.,  2013, \mn@doi [\mnras]
  {10.1093/mnras/stt723}, \href
  {http://adsabs.harvard.edu/abs/2013MNRAS.432.2647H} {432, 2647}

\bibitem[\protect\citeauthoryear{{Hopkins}, {Kere{\v s}}, {O{\~n}orbe},
  {Faucher-Gigu{\`e}re}, {Quataert}, {Murray}  \& {Bullock}}{{Hopkins}
  et~al.}{2014}]{FIRE}
{Hopkins} P.~F.,  {Kere{\v s}} D.,  {O{\~n}orbe} J.,  {Faucher-Gigu{\`e}re}
  C.-A.,  {Quataert} E.,  {Murray} N.,   {Bullock} J.~S.,  2014, \mn@doi
  [\mnras] {10.1093/mnras/stu1738}, \href
  {http://adsabs.harvard.edu/abs/2014MNRAS.445..581H} {445, 581}

\bibitem[\protect\citeauthoryear{{Hopkins} et~al.,}{{Hopkins}
  et~al.}{2018}]{FIRE2}
{Hopkins} P.~F.,  et~al., 2018, \mn@doi [\mnras] {10.1093/mnras/sty1690}, \href
  {http://adsabs.harvard.edu/abs/2018MNRAS.480..800H} {480, 800}

\bibitem[\protect\citeauthoryear{Hunter}{Hunter}{2007}]{Matplotlib}
Hunter J.~D.,  2007, Computing In Science \& Engineering, 9, 90

\bibitem[\protect\citeauthoryear{Jones, Oliphant, Peterson  et~al.}{Jones
  et~al.}{01  }]{scipy}
Jones E.,  Oliphant T.,  Peterson P.,   et~al., 2001--, {SciPy}: Open source
  scientific tools for {Python}, \url {http://www.scipy.org/}

\bibitem[\protect\citeauthoryear{{Katz} \& {White}}{{Katz} \&
  {White}}{1993}]{Katz1993}
{Katz} N.,  {White} S.~D.~M.,  1993, \mn@doi [\apj] {10.1086/172935}, \href
  {http://adsabs.harvard.edu/abs/1993ApJ...412..455K} {412, 455}

\bibitem[\protect\citeauthoryear{{Kawata} \& {Mulchaey}}{{Kawata} \&
  {Mulchaey}}{2008}]{Kawata2008}
{Kawata} D.,  {Mulchaey} J.~S.,  2008, \mn@doi [\apjl] {10.1086/526544}, \href
  {http://adsabs.harvard.edu/abs/2008ApJ...672L.103K} {672, L103}

\bibitem[\protect\citeauthoryear{{Knollmann} \& {Knebe}}{{Knollmann} \&
  {Knebe}}{2011}]{AHF}
{Knollmann} S.~R.,  {Knebe} A.,  2011, {AHF: Amiga's Halo Finder}, Astrophysics
  Source Code Library (\mn@eprint {ascl} {1102.009})

\bibitem[\protect\citeauthoryear{{Komatsu} et~al.,}{{Komatsu}
  et~al.}{2011}]{Komatsu2011}
{Komatsu} E.,  et~al., 2011, \mn@doi [\apjs] {10.1088/0067-0049/192/2/18},
  \href {http://adsabs.harvard.edu/abs/2011ApJS..192...18K} {192, 18}

\bibitem[\protect\citeauthoryear{{Kroupa}}{{Kroupa}}{2001}]{Kroupa2001}
{Kroupa} P.,  2001, \mn@doi [\mnras] {10.1046/j.1365-8711.2001.04022.x}, \href
  {http://adsabs.harvard.edu/abs/2001MNRAS.322..231K} {322, 231}

\bibitem[\protect\citeauthoryear{{Krumholz} \& {Gnedin}}{{Krumholz} \&
  {Gnedin}}{2011}]{Krumholz2011}
{Krumholz} M.~R.,  {Gnedin} N.~Y.,  2011, \mn@doi [\apj]
  {10.1088/0004-637X/729/1/36}, \href
  {http://adsabs.harvard.edu/abs/2011ApJ...729...36K} {729, 36}

\bibitem[\protect\citeauthoryear{{Larson}, {Tinsley}  \& {Caldwell}}{{Larson}
  et~al.}{1980}]{Larson1980}
{Larson} R.~B.,  {Tinsley} B.~M.,   {Caldwell} C.~N.,  1980, \mn@doi [\apj]
  {10.1086/157917}, \href {http://adsabs.harvard.edu/abs/1980ApJ...237..692L}
  {237, 692}

\bibitem[\protect\citeauthoryear{{Larson} et~al.,}{{Larson}
  et~al.}{2011}]{Larson2011}
{Larson} D.,  et~al., 2011, \mn@doi [\apjs] {10.1088/0067-0049/192/2/16}, \href
  {http://adsabs.harvard.edu/abs/2011ApJS..192...16L} {192, 16}

\bibitem[\protect\citeauthoryear{{Lovell} et~al.,}{{Lovell}
  et~al.}{2017}]{Lovell2017}
{Lovell} M.~R.,  et~al., 2017, \mn@doi [\mnras] {10.1093/mnras/stx654}, \href
  {http://adsabs.harvard.edu/abs/2017MNRAS.468.4285L} {468, 4285}

\bibitem[\protect\citeauthoryear{{Lunnan}, {Vogelsberger}, {Frebel},
  {Hernquist}, {Lidz}  \& {Boylan-Kolchin}}{{Lunnan} et~al.}{2012}]{Lunnan2012}
{Lunnan} R.,  {Vogelsberger} M.,  {Frebel} A.,  {Hernquist} L.,  {Lidz} A.,
  {Boylan-Kolchin} M.,  2012, \mn@doi [\apj] {10.1088/0004-637X/746/1/109},
  \href {http://adsabs.harvard.edu/abs/2012ApJ...746..109L} {746, 109}

\bibitem[\protect\citeauthoryear{{McConnachie}}{{McConnachie}}{2012}]{McConnachie2012}
{McConnachie} A.~W.,  2012, \mn@doi [\aj] {10.1088/0004-6256/144/1/4}, \href
  {http://adsabs.harvard.edu/abs/2012AJ....144....4M} {144, 4}

\bibitem[\protect\citeauthoryear{{Necib}, {Lisanti}, {Garrison-Kimmel},
  {Wetzel}, {Sanderson}, {Hopkins}, {Faucher-Gigu{\`e}re}  \& {Kere{\v
  s}}}{{Necib} et~al.}{2018}]{Necib2018}
{Necib} L.,  {Lisanti} M.,  {Garrison-Kimmel} S.,  {Wetzel} A.,  {Sanderson}
  R.,  {Hopkins} P.~F.,  {Faucher-Gigu{\`e}re} C.-A.,   {Kere{\v s}} D.,  2018,
  preprint, \href {http://adsabs.harvard.edu/abs/2018arXiv181012301N} {}
  (\mn@eprint {arXiv} {1810.12301})

\bibitem[\protect\citeauthoryear{{Norman}, {Chen}, {Wise}  \& {Xu}}{{Norman}
  et~al.}{2018}]{Chen2017}
{Norman} M.~L.,  {Chen} P.,  {Wise} J.~H.,   {Xu} H.,  2018, \mn@doi [\apj]
  {10.3847/1538-4357/aae30b}, \href
  {https://ui.adsabs.harvard.edu/\#abs/2018ApJ...867...27N} {867, 27}

\bibitem[\protect\citeauthoryear{{Nulsen}}{{Nulsen}}{1982}]{Nulsen1982}
{Nulsen} P.~E.~J.,  1982, \mn@doi [\mnras] {10.1093/mnras/198.4.1007}, \href
  {http://adsabs.harvard.edu/abs/1982MNRAS.198.1007N} {198, 1007}

\bibitem[\protect\citeauthoryear{{O{\~n}orbe}, {Garrison-Kimmel}, {Maller},
  {Bullock}, {Rocha}  \& {Hahn}}{{O{\~n}orbe} et~al.}{2014}]{Onorbe2014}
{O{\~n}orbe} J.,  {Garrison-Kimmel} S.,  {Maller} A.~H.,  {Bullock} J.~S.,
  {Rocha} M.,   {Hahn} O.,  2014, \mn@doi [\mnras] {10.1093/mnras/stt2020},
  \href {http://adsabs.harvard.edu/abs/2014MNRAS.437.1894O} {437, 1894}

\bibitem[\protect\citeauthoryear{{O{\~n}orbe}, {Boylan-Kolchin}, {Bullock},
  {Hopkins}, {Kere{\v s}}, {Faucher-Gigu{\`e}re}, {Quataert}  \&
  {Murray}}{{O{\~n}orbe} et~al.}{2015}]{Onorbe2015}
{O{\~n}orbe} J.,  {Boylan-Kolchin} M.,  {Bullock} J.~S.,  {Hopkins} P.~F.,
  {Kere{\v s}} D.,  {Faucher-Gigu{\`e}re} C.-A.,  {Quataert} E.,   {Murray} N.,
   2015, \mn@doi [\mnras] {10.1093/mnras/stv2072}, \href
  {http://adsabs.harvard.edu/abs/2015MNRAS.454.2092O} {454, 2092}

\bibitem[\protect\citeauthoryear{{Ocvirk} et~al.,}{{Ocvirk}
  et~al.}{2016}]{Ocvirk2016}
{Ocvirk} P.,  et~al., 2016, \mn@doi [\mnras] {10.1093/mnras/stw2036}, \href
  {http://adsabs.harvard.edu/abs/2016MNRAS.463.1462O} {463, 1462}

\bibitem[\protect\citeauthoryear{{Pearson} et~al.,}{{Pearson}
  et~al.}{2016}]{Pearson2016}
{Pearson} S.,  et~al., 2016, \mn@doi [\mnras] {10.1093/mnras/stw757}, \href
  {http://adsabs.harvard.edu/abs/2016MNRAS.459.1827P} {459, 1827}

\bibitem[\protect\citeauthoryear{{Pentericci} et~al.,}{{Pentericci}
  et~al.}{2014}]{Pentericci2014}
{Pentericci} L.,  et~al., 2014, \mn@doi [\apj] {10.1088/0004-637X/793/2/113},
  \href {https://ui.adsabs.harvard.edu/#abs/2014ApJ...793..113P} {793, 113}

\bibitem[\protect\citeauthoryear{Perez \& Granger}{Perez \&
  Granger}{2007}]{ipython}
Perez F.,  Granger B.~E.,  2007, \mn@doi [Computing in Science Engineering]
  {10.1109/MCSE.2007.53}, 9, 21

\bibitem[\protect\citeauthoryear{Pettitt}{Pettitt}{1976}]{Pettitt1976:2SampADTest}
Pettitt A.~N.,  1976, Biometrika, 63, 161

\bibitem[\protect\citeauthoryear{{Planck Collaboration} et~al.,}{{Planck
  Collaboration} et~al.}{2016}]{Planck15}
{Planck Collaboration} et~al., 2016, \mn@doi [\aap]
  {10.1051/0004-6361/201525830}, \href
  {http://adsabs.harvard.edu/abs/2016A%26A...594A..13P} {594, A13}

\bibitem[\protect\citeauthoryear{{Planck Collaboration} et~al.,}{{Planck
  Collaboration} et~al.}{2018}]{Planck2018}
{Planck Collaboration} et~al., 2018, preprint, \href
  {http://adsabs.harvard.edu/abs/2018arXiv180706209P} {} (\mn@eprint {arXiv}
  {1807.06209})

\bibitem[\protect\citeauthoryear{{Read}, {Agertz}  \& {Collins}}{{Read}
  et~al.}{2016}]{Read2016}
{Read} J.~I.,  {Agertz} O.,   {Collins} M.~L.~M.,  2016, \mn@doi [\mnras]
  {10.1093/mnras/stw713}, \href
  {http://adsabs.harvard.edu/abs/2016MNRAS.459.2573R} {459, 2573}

\bibitem[\protect\citeauthoryear{{Read}, {Walker}  \& {Steger}}{{Read}
  et~al.}{2019}]{Read2018}
{Read} J.~I.,  {Walker} M.~G.,   {Steger} P.,  2019, \mn@doi [\mnras]
  {10.1093/mnras/sty3404}, \href
  {https://ui.adsabs.harvard.edu/\#abs/2019MNRAS.484.1401R} {484, 1401}

\bibitem[\protect\citeauthoryear{{Ricotti}, {Gnedin}  \& {Shull}}{{Ricotti}
  et~al.}{2008}]{Ricotti2008}
{Ricotti} M.,  {Gnedin} N.~Y.,   {Shull} J.~M.,  2008, \mn@doi [\apj]
  {10.1086/590901}, \href {http://adsabs.harvard.edu/abs/2008ApJ...685...21R}
  {685, 21}

\bibitem[\protect\citeauthoryear{{Rocha}, {Peter}  \& {Bullock}}{{Rocha}
  et~al.}{2012}]{Rocha2012}
{Rocha} M.,  {Peter} A.~H.~G.,   {Bullock} J.,  2012, \mn@doi [\mnras]
  {10.1111/j.1365-2966.2012.21432.x}, \href
  {http://adsabs.harvard.edu/abs/2012MNRAS.425..231R} {425, 231}

\bibitem[\protect\citeauthoryear{{Rodriguez-Gomez} et~al.,}{{Rodriguez-Gomez}
  et~al.}{2016}]{RodriguezGomez2016}
{Rodriguez-Gomez} V.,  et~al., 2016, \mn@doi [\mnras] {10.1093/mnras/stw456},
  \href {http://adsabs.harvard.edu/abs/2016MNRAS.458.2371R} {458, 2371}

\bibitem[\protect\citeauthoryear{{Rodriguez Wimberly}, {Cooper}, {Fillingham},
  {Boylan-Kolchin}, {Bullock}  \& {Garrison- Kimmel}}{{Rodriguez Wimberly}
  et~al.}{2019}]{RodriguezWimberly2018}
{Rodriguez Wimberly} M.~K.,  {Cooper} M.~C.,  {Fillingham} S.~P.,
  {Boylan-Kolchin} M.,  {Bullock} J.~S.,   {Garrison- Kimmel} S.,  2019,
  \mn@doi [\mnras] {10.1093/mnras/sty3357}, \href
  {https://ui.adsabs.harvard.edu/\#abs/2019MNRAS.483.4031R} {483, 4031}

\bibitem[\protect\citeauthoryear{{Sanderson} et~al.,}{{Sanderson}
  et~al.}{2018}]{Sanderson2017}
{Sanderson} R.~E.,  et~al., 2018, \mn@doi [\apj] {10.3847/1538-4357/aaeb33},
  \href {https://ui.adsabs.harvard.edu/\#abs/2018ApJ...869...12S} {869, 12}

\bibitem[\protect\citeauthoryear{{Sawala} et~al.,}{{Sawala}
  et~al.}{2016}]{Sawala2016}
{Sawala} T.,  et~al., 2016, \mn@doi [\mnras] {10.1093/mnras/stw145}, \href
  {http://adsabs.harvard.edu/abs/2016MNRAS.457.1931S} {457, 1931}

\bibitem[\protect\citeauthoryear{Scholz \& Stephens}{Scholz \&
  Stephens}{1987}]{Scholz1987:kSampADTest}
Scholz F.~W.,  Stephens M.~A.,  1987, \mn@doi [Journal of the American
  Statistical Association] {10.1080/01621459.1987.10478517}, 82, 918

\bibitem[\protect\citeauthoryear{{Simpson}, {Grand}, {G{\'o}mez}, {Marinacci},
  {Pakmor}, {Springel}, {Campbell}  \& {Frenk}}{{Simpson}
  et~al.}{2018}]{Simpson2018}
{Simpson} C.~M.,  {Grand} R.~J.~J.,  {G{\'o}mez} F.~A.,  {Marinacci} F.,
  {Pakmor} R.,  {Springel} V.,  {Campbell} D.~J.~R.,   {Frenk} C.~S.,  2018,
  \mn@doi [\mnras] {10.1093/mnras/sty774}, \href
  {http://adsabs.harvard.edu/abs/2018MNRAS.478..548S} {478, 548}

\bibitem[\protect\citeauthoryear{{Skillman} et~al.,}{{Skillman}
  et~al.}{2017}]{Skillman2017}
{Skillman} E.~D.,  et~al., 2017, \mn@doi [\apj] {10.3847/1538-4357/aa60c5},
  \href {http://adsabs.harvard.edu/abs/2017ApJ...837..102S} {837, 102}

\bibitem[\protect\citeauthoryear{Smith \& Lang}{Smith \& Lang}{2018}]{ytree}
Smith B.,  Lang M.,  2018, ytree: merger-tree toolkit,
  \mn@doi{10.5281/zenodo.1174374}, \url
  {https://doi.org/10.5281/zenodo.1174374}

\bibitem[\protect\citeauthoryear{{Spekkens}, {Urbancic}, {Mason}, {Willman}  \&
  {Aguirre}}{{Spekkens} et~al.}{2014}]{Spekkens2014}
{Spekkens} K.,  {Urbancic} N.,  {Mason} B.~S.,  {Willman} B.,   {Aguirre}
  J.~E.,  2014, \mn@doi [\apjl] {10.1088/2041-8205/795/1/L5}, \href
  {http://adsabs.harvard.edu/abs/2014ApJ...795L...5S} {795, L5}

\bibitem[\protect\citeauthoryear{{Su}, {Hopkins}, {Hayward},
  {Faucher-Gigu{\`e}re}, {Kere{\v s}}, {Ma}  \& {Robles}}{{Su}
  et~al.}{2017}]{Su2016}
{Su} K.-Y.,  {Hopkins} P.~F.,  {Hayward} C.~C.,  {Faucher-Gigu{\`e}re} C.-A.,
  {Kere{\v s}} D.,  {Ma} X.,   {Robles} V.~H.,  2017, \mn@doi [\mnras]
  {10.1093/mnras/stx1463}, \href
  {http://adsabs.harvard.edu/abs/2017MNRAS.471..144S} {471, 144}

\bibitem[\protect\citeauthoryear{{Su} et~al.,}{{Su} et~al.}{2018}]{Su2018}
{Su} K.-Y.,  et~al., 2018, \mn@doi [\mnras] {10.1093/mnras/sty1928}, \href
  {http://adsabs.harvard.edu/abs/2018MNRAS.480.1666S} {480, 1666}

\bibitem[\protect\citeauthoryear{{Tanaka}, {Chiba}, {Hayashi}, {Komiyama},
  {Okamoto}, {Cooper}, {Okamoto}  \& {Spitler}}{{Tanaka}
  et~al.}{2018}]{Tanaka2018}
{Tanaka} M.,  {Chiba} M.,  {Hayashi} K.,  {Komiyama} Y.,  {Okamoto} T.,
  {Cooper} A.~P.,  {Okamoto} S.,   {Spitler} L.,  2018, \mn@doi [\apj]
  {10.3847/1538-4357/aad9fe}, \href
  {https://ui.adsabs.harvard.edu/\#abs/2018ApJ...865..125T} {865, 125}

\bibitem[\protect\citeauthoryear{{Teyssier}, {Johnston}  \&
  {Kuhlen}}{{Teyssier} et~al.}{2012}]{Teyssier2012}
{Teyssier} M.,  {Johnston} K.~V.,   {Kuhlen} M.,  2012, \mn@doi [\mnras]
  {10.1111/j.1365-2966.2012.21793.x}, \href
  {http://adsabs.harvard.edu/abs/2012MNRAS.426.1808T} {426, 1808}

\bibitem[\protect\citeauthoryear{{The Astropy Collaboration} et~al.,}{{The
  Astropy Collaboration} et~al.}{2018}]{Astropy2}
{The Astropy Collaboration} et~al., 2018, \mn@doi [\aj]
  {10.3847/1538-3881/aabc4f}, \href
  {http://adsabs.harvard.edu/abs/2018AJ....156..123T} {156, 123}

\bibitem[\protect\citeauthoryear{{Trac} \& {Cen}}{{Trac} \&
  {Cen}}{2007}]{Trac2007}
{Trac} H.,  {Cen} R.,  2007, \mn@doi [\apj] {10.1086/522566}, \href
  {http://adsabs.harvard.edu/abs/2007ApJ...671....1T} {671, 1}

\bibitem[\protect\citeauthoryear{{Turk}, {Smith}, {Oishi}, {Skory}, {Skillman},
  {Abel}  \& {Norman}}{{Turk} et~al.}{2011}]{yt}
{Turk} M.~J.,  {Smith} B.~D.,  {Oishi} J.~S.,  {Skory} S.,  {Skillman} S.~W.,
  {Abel} T.,   {Norman} M.~L.,  2011, \mn@doi [The Astrophysical Journal
  Supplement Series] {10.1088/0067-0049/192/1/9}, \href
  {http://adsabs.harvard.edu/abs/2011ApJS..192....9T} {192, 9}

\bibitem[\protect\citeauthoryear{{Wadsley}, {Stadel}  \& {Quinn}}{{Wadsley}
  et~al.}{2004}]{Wadsley2004}
{Wadsley} J.~W.,  {Stadel} J.,   {Quinn} T.,  2004, \mn@doi [\na]
  {10.1016/j.newast.2003.08.004}, \href
  {http://adsabs.harvard.edu/abs/2004NewA....9..137W} {9, 137}

\bibitem[\protect\citeauthoryear{{Weisz}, {Dolphin}, {Skillman}, {Holtzman},
  {Gilbert}, {Dalcanton}  \& {Williams}}{{Weisz} et~al.}{2014a}]{WeiszSFH}
{Weisz} D.~R.,  {Dolphin} A.~E.,  {Skillman} E.~D.,  {Holtzman} J.,  {Gilbert}
  K.~M.,  {Dalcanton} J.~J.,   {Williams} B.~F.,  2014a, \mn@doi [\apj]
  {10.1088/0004-637X/789/2/147}, \href
  {http://adsabs.harvard.edu/abs/2014ApJ...789..147W} {789, 147}

\bibitem[\protect\citeauthoryear{{Weisz}, {Dolphin}, {Skillman}, {Holtzman},
  {Gilbert}, {Dalcanton}  \& {Williams}}{{Weisz}
  et~al.}{2014b}]{Weisz2014:ReionSigs}
{Weisz} D.~R.,  {Dolphin} A.~E.,  {Skillman} E.~D.,  {Holtzman} J.,  {Gilbert}
  K.~M.,  {Dalcanton} J.~J.,   {Williams} B.~F.,  2014b, \mn@doi [\apj]
  {10.1088/0004-637X/789/2/148}, \href
  {http://adsabs.harvard.edu/abs/2014ApJ...789..148W} {789, 148}

\bibitem[\protect\citeauthoryear{{Weisz}, {Johnson}  \& {Conroy}}{{Weisz}
  et~al.}{2014c}]{Weisz2014b}
{Weisz} D.~R.,  {Johnson} B.~D.,   {Conroy} C.,  2014c, \mn@doi [\apj]
  {10.1088/2041-8205/794/1/L3}, \href
  {https://ui.adsabs.harvard.edu/#abs/2014ApJ...794L...3W} {794, L3}

\bibitem[\protect\citeauthoryear{{Weisz}, {Dolphin}, {Skillman}, {Holtzman},
  {Gilbert}, {Dalcanton}  \& {Williams}}{{Weisz} et~al.}{2015}]{Weisz2015}
{Weisz} D.~R.,  {Dolphin} A.~E.,  {Skillman} E.~D.,  {Holtzman} J.,  {Gilbert}
  K.~M.,  {Dalcanton} J.~J.,   {Williams} B.~F.,  2015, \mn@doi [\apj]
  {10.1088/0004-637X/804/2/136}, \href
  {http://adsabs.harvard.edu/abs/2015ApJ...804..136W} {804, 136}

\bibitem[\protect\citeauthoryear{{Wetzel}, {Tinker}, {Conroy}  \& {van den
  Bosch}}{{Wetzel} et~al.}{2013}]{Wetzel2013}
{Wetzel} A.~R.,  {Tinker} J.~L.,  {Conroy} C.,   {van den Bosch} F.~C.,  2013,
  \mn@doi [\mnras] {10.1093/mnras/stt469}, \href
  {http://adsabs.harvard.edu/abs/2013MNRAS.432..336W} {432, 336}

\bibitem[\protect\citeauthoryear{{Wetzel}, {Tollerud}  \& {Weisz}}{{Wetzel}
  et~al.}{2015}]{Wetzel2015b}
{Wetzel} A.~R.,  {Tollerud} E.~J.,   {Weisz} D.~R.,  2015, \mn@doi [\apjl]
  {10.1088/2041-8205/808/1/L27}, \href
  {http://adsabs.harvard.edu/abs/2015ApJ...808L..27W} {808, L27}

\bibitem[\protect\citeauthoryear{{Wetzel}, {Hopkins}, {Kim},
  {Faucher-Gigu{\`e}re}, {Kere{\v s}}  \& {Quataert}}{{Wetzel}
  et~al.}{2016}]{Wetzel2016}
{Wetzel} A.~R.,  {Hopkins} P.~F.,  {Kim} J.-h.,  {Faucher-Gigu{\`e}re} C.-A.,
  {Kere{\v s}} D.,   {Quataert} E.,  2016, \mn@doi [\apjl]
  {10.3847/2041-8205/827/2/L23}, \href
  {http://adsabs.harvard.edu/abs/2016ApJ...827L..23W} {827, L23}

\bibitem[\protect\citeauthoryear{{Wheeler}, {Phillips}, {Cooper},
  {Boylan-Kolchin}  \& {Bullock}}{{Wheeler} et~al.}{2014}]{Wheeler2014}
{Wheeler} C.,  {Phillips} J.~I.,  {Cooper} M.~C.,  {Boylan-Kolchin} M.,
  {Bullock} J.~S.,  2014, \mn@doi [\mnras] {10.1093/mnras/stu965}, \href
  {http://adsabs.harvard.edu/abs/2014MNRAS.442.1396W} {442, 1396}

\bibitem[\protect\citeauthoryear{{Wheeler}, {O{\~n}orbe}, {Bullock},
  {Boylan-Kolchin}, {Elbert}, {Garrison- Kimmel}, {Hopkins}  \&
  {Kere{\v{s}}}}{{Wheeler} et~al.}{2015}]{Wheeler2015}
{Wheeler} C.,  {O{\~n}orbe} J.,  {Bullock} J.~S.,  {Boylan-Kolchin} M.,
  {Elbert} O.~D.,  {Garrison- Kimmel} S.,  {Hopkins} P.~F.,   {Kere{\v{s}}} D.,
   2015, \mn@doi [\mnras] {10.1093/mnras/stv1691}, \href
  {https://ui.adsabs.harvard.edu/\#abs/2015MNRAS.453.1305W} {453, 1305}

\bibitem[\protect\citeauthoryear{{Wheeler} et~al.,}{{Wheeler}
  et~al.}{2018}]{Wheeler2018}
{Wheeler} C.,  et~al., 2018, preprint, \href
  {http://adsabs.harvard.edu/abs/2018arXiv181202749W} {} (\mn@eprint {arXiv}
  {1812.02749})

\bibitem[\protect\citeauthoryear{{Wright}, {Brooks}, {Weisz}  \&
  {Christensen}}{{Wright} et~al.}{2019}]{Wright2018}
{Wright} A.~C.,  {Brooks} A.~M.,  {Weisz} D.~R.,   {Christensen} C.~R.,  2019,
  \mn@doi [\mnras] {10.1093/mnras/sty2759}, \href
  {https://ui.adsabs.harvard.edu/\#abs/2019MNRAS.482.1176W} {482, 1176}

\bibitem[\protect\citeauthoryear{{Zheng} et~al.,}{{Zheng}
  et~al.}{2017}]{Zheng2017}
{Zheng} Z.-Y.,  et~al., 2017, \mn@doi [\apjl] {10.3847/2041-8213/aa794f}, \href
  {http://adsabs.harvard.edu/abs/2017ApJ...842L..22Z} {842, L22}

\bibitem[\protect\citeauthoryear{van~der Walt, Colbert  \& Varoquaux}{van~der
  Walt et~al.}{2011}]{numpy}
van~der Walt S.,  Colbert S.~C.,   Varoquaux G.,  2011, \mn@doi [Computing in
  Science Engineering] {10.1109/MCSE.2011.37}, 13, 22

\makeatother
\end{thebibliography}

\appendix

\section{Resolution dependence}
\label{sec:resolution}

\S\ref{ssec:caveats} discusses the expected impact of resolution
on our results.  We justify the conclusions we reach in that Section 
here, but also refer the reader to \S~4.1.3 of \citet{FIRE2} for a 
longer discussion of the resolution elements required for the 
convergence of various galactic properties in the FIRE-2 simulations.

\begin{figure}
    \centering
    \includegraphics[width=\columnwidth]{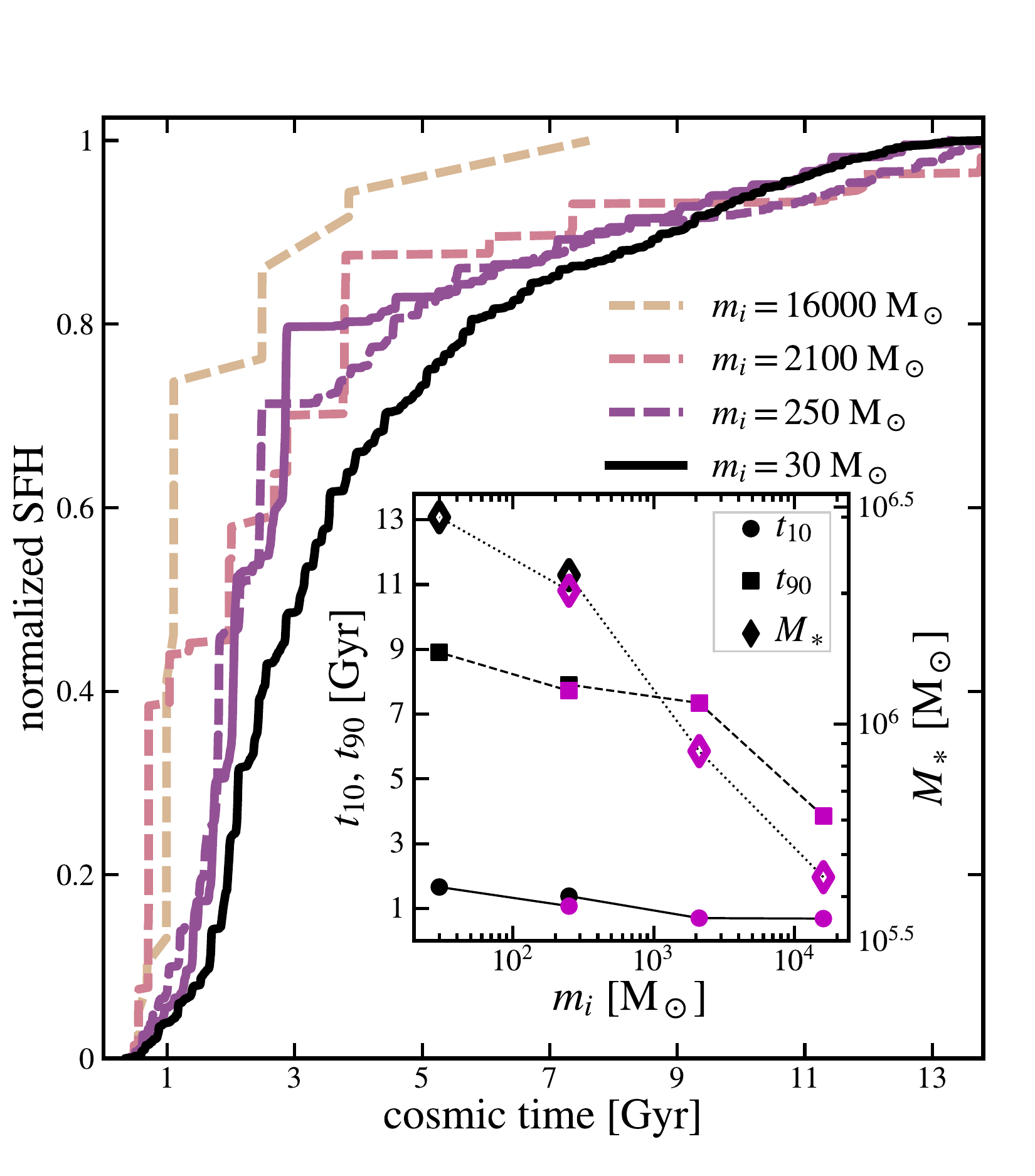}
    \caption{The SFH of a single isolated dwarf central galaxy (\run{m10q},
    with $\mstar\simeq10^6\msun$) in runs with initial gas particle masses 
    $m_i = 30~\ - 16,000\msun$.  For comparison, the LG and isolated MW 
    simulations have resolutions $m_i = 3500$--$7100\msun$.  The inset panel 
    plots $\tten$ and $\tnintey$ for each run (squares and circles, left axis), 
    as well as the stellar mass of the galaxy formed in each run (diamonds, 
    right axis).  Simulations that do \emph{not} include subgrid metal diffusion 
    are indicated with dashed lines and magenta symbols.  \run{m10q} was simulated 
    with $m_i = 250\msun$ both with and without this subgrid prescription; the SFHs 
    from these two runs are nearly identical, indicating that metal diffusion has 
    a negligible impact on the shape of the SFH.  At lower resolutions (higher $m_i$), 
    the galaxy forms stars earlier.  However, the $m_i=2100\msun$ run agrees well 
    with the run with $m_i=250\msun$, with strongly divergent behavior only appearing 
    for $m_i = 16,000\msun$ (more than twice the lowest resolution isolated-MW 
    simulation).  Moreover, these changes are also in line with the $\mstar$ 
    trends discussed in the main text:  at lower resolution, the galaxy is also 
    lower mass.}
    \label{fig:m10qres}
\end{figure}

We begin with Figure~\ref{fig:m10qres}, which presents the SFH of a 
\emph{single}, low mass, isolated dwarf galaxy (\run{m10q}) as 
simulated at variety of resolutions.  As we decrease the resolution 
(quantified by the initial gas particle mass $m_i$) from the ultra-high 
resolution $m_i = 30\msun$ simulation that we analyze in the main text, 
the galaxy tends to form a greater fraction of its stars at 
later times.  At the lowest resolution -- which is even lower
resolution than the simulations targeting isolated MW-mass hosts in 
the main text -- the dwarf even artificially quenches at $t\simeq7~\gyr$.  
However, the SFH is consistent between $m_i=250\msun$ and $m_i=2100\msun$, 
suggesting that the critical resolution at which we stop resolving
the SFH of this dwarf (with $\mstar\simeq2\times10^6\msun$) occurs 
between $m_i=2100$--$16,000\msun$.  Unfortunately, our simulations
that include MW-mass host(s) lie in between these two resolutions,
though they do fall slightly closer to the lower edge of that range.
Therefore, we acknowledge that our galaxies with $\mstar\lesssim10^6\msun$
are somewhat under-resolved and likely should have a higher fraction 
of late time star formation.  As we argue in \S\ref{ssec:caveats},
however, correcting for resolution effects should act to amplify 
the differences between the LG-like simulations and those of isolated 
MW-mass hosts, as the dwarf galaxies in the (slightly) higher resolution 
LG-like simulations already appear to have reached a given fraction of
their $z=0$ stellar mass earlier.  Moreover, the changes in 
Figure~\ref{fig:m10qres} with resolution are also consistent with the 
mass trends identified in the main  text:  as $m_i$ increases (resolution 
decreases), the galaxy forms earlier, but also reaches a smaller overall 
stellar mass.

\begin{figure}
    \centering
    \includegraphics[width=\columnwidth]{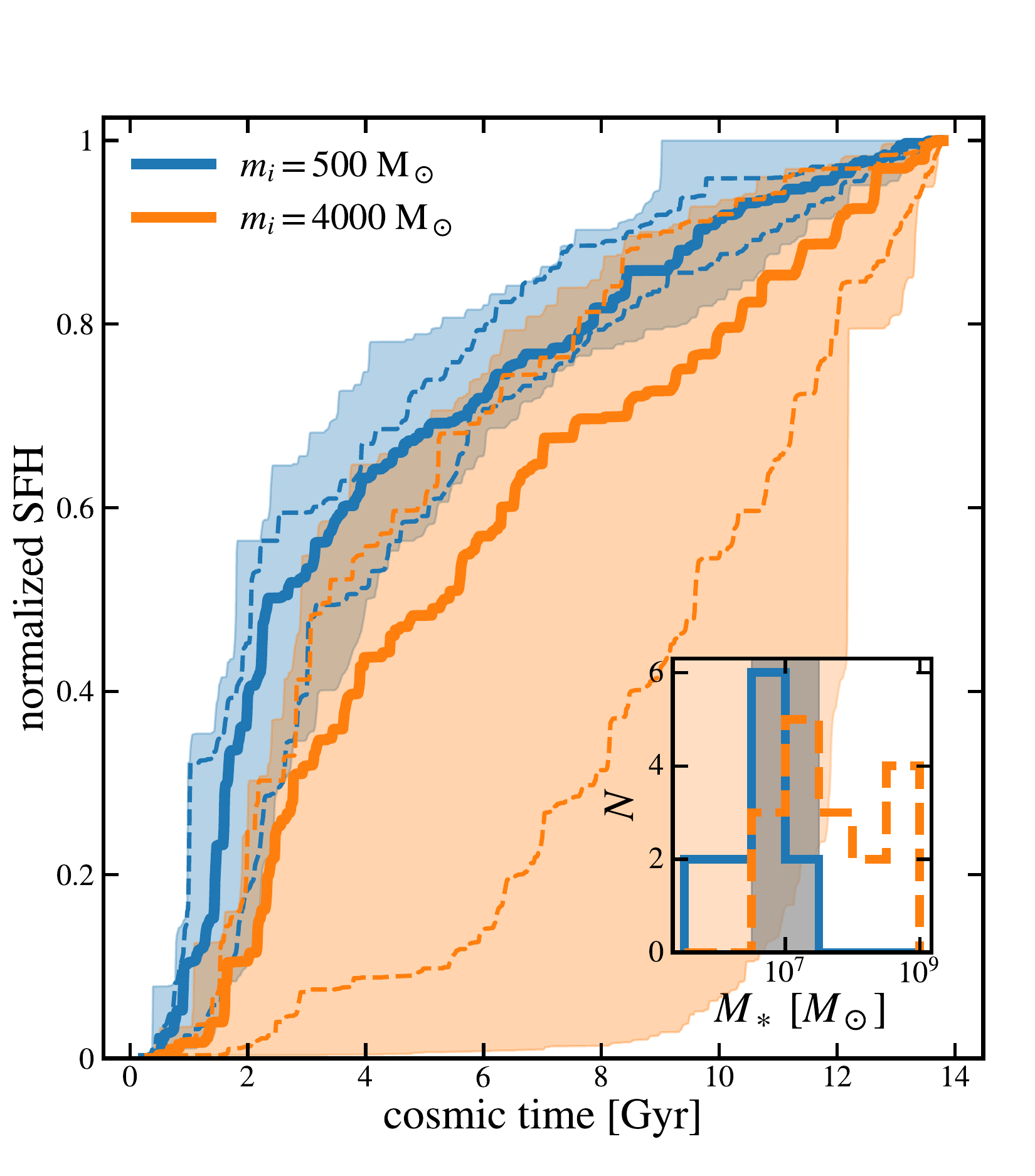}
    \caption{SFHs of highly isolated dwarf galaxies with stellar masses 
    $\mstar=10^{6.5}-10^{7.5}\msun$; i.e., using a subset of the 
    \citealp{Fitts2017} and \citet{Graus2019} samples.  
    Dashed lines indicate 68\% of the sample, and the shaded areas 
    indicate the full spreads.  The lower resolution simulations tend to 
    quench slightly later, in the opposite direction from the resolution 
    trends in Figure~\ref{fig:m10qres}, but we point out from the inset 
    panel that the lower resolution sample also tends towards higher masses, 
    so the difference is in line with the stellar mass trends identified
    in the main text.}
    \label{fig:isolatedRes}
\end{figure}

Figure~\ref{fig:isolatedRes} demonstrates that these mass trends,
together with galaxy-by-galaxy variances, dominate over resolution 
effects (at our resolution).  Here we select highly isolated dwarf
central galaxies with $\mstar=10^{6.5}$--$10^{7.5}\msun$ from two 
different sets of galaxies: the \citet[][in blue]{Fitts2017} and 
\citet[][in orange]{Graus2019} samples.  That is, the two sets plotted 
in Figure~\ref{fig:isolatedRes} are \emph{not} the same galaxies 
at two different resolutions (unlike Figure~\ref{fig:m10qres}).
Though the \citet{Graus2019} sample is at lower resolution, those 
galaxies form later than the dwarfs in the higher 
resolution \citet{Fitts2017} sample.  As indicated by the inset panel, 
which plots the stellar masses of the two samples (with the shaded 
region representing the $\mstar$ range of galaxies plotted in the 
main Figure), the latter sample is at lower average mass (within this 
1~dex mass bin). 

\begin{figure}
    \centering
    \includegraphics[width=\columnwidth]{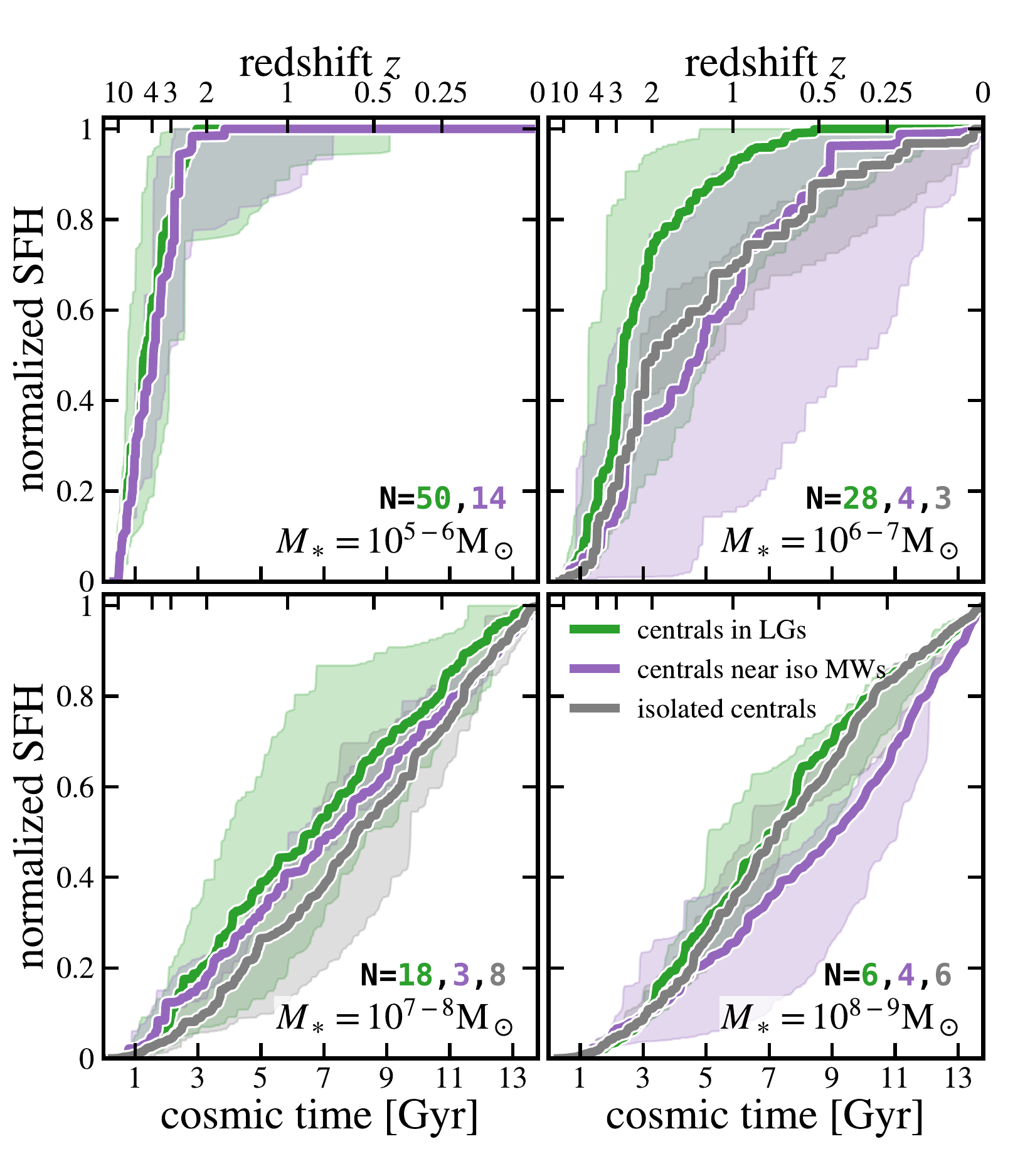}
    \caption{Comparing the star formation histories of dwarf 
    centrals from different environments at roughly fixed 
    resolution ($m_i = 3500$--$4200\msun$).  Though the 
    statistics are relatively poor in any individual bin (other 
    than around the LGs), the conclusions identified above 
    roughly hold true:  dwarf centrals in LG-like environments 
    form earlier than both the highly isolated centrals and those 
    around a single MW.}
    \label{fig:sameres-centrals}
\end{figure}

As a final check on the impact of resolution, Figure~\ref{fig:sameres-centrals}
plots the SFHs of dwarf centrals from simulations that have approximately
identical resolution:  the LG-like pairs ($m_i = 3,500$~and~$4,000\msun$),
the \citet{Graus2019} sample ($m_i = 4,000\msun$), and \run{m12z} ($m_i=4,200\msun$).
Though the sample size is far too small to draw meaningful conclusions 
overall, the results identified above hold even in this limited set of 
simulations.

\section{Effects of a later reionization} 
\label{sec:reion_effects}

\begin{figure}
    \centering
    \includegraphics[width=\columnwidth]{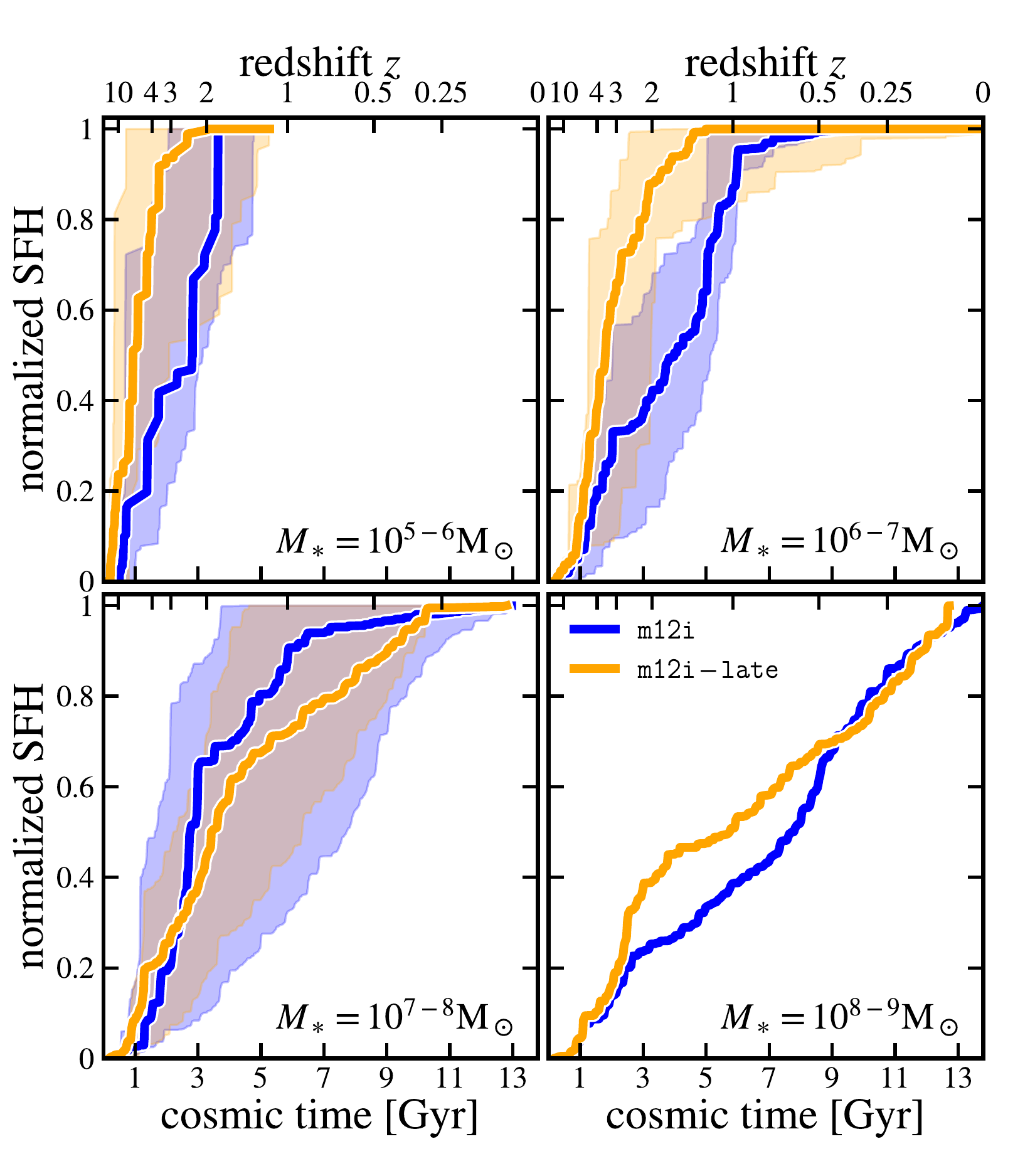}
    \caption{Comparing the SFHs of satellites around a single MW-mass
    host (\run{m12i}) run with the standard reionizing background (blue) 
    and with a version of the \citet{FaucherGiguere2009} UV background modified to produce a later reionization history that is consistent with the \citep{Planck2018} electron scattering optical depth (orange). In this model, the hydrogen neutral fraction drops to $0.5$ at $z\sim7.8$. The thick lines take 
    the median in each mass bin and the shaded regions indicate the full 
    extent of the scatter.  As discussed in the text, an earlier $z_\mathrm{reion}$ 
    tends to reduce the relative amount of early time star formation, 
    as there is less time for stars to form before the background begins 
    to play a role; the lines therefore tend to shift slightly to the 
    left (more stars formed at earlier times) when moving to a background
    that reionizes the Universe at a later time.
    \label{lastpage}}
    \label{fig:m12i-sats-reion}
\end{figure}

As discussed in \S\ref{ssec:caveats}, all of the simulations analyzed 
in the main text adopt a reionizing background that completes reionization
at $z\simeq10$, while more recent observations suggest the Universe 
reionized quickly at $z\sim7$.  Figure~\ref{fig:m12i-sats-reion}
demonstrates how changing $\zreion$ alters the SFHs of the satellites
around a single (isolated) MW-mass host, \run{m12i}.  As described in
\S\ref{ssec:caveats}, a later $\zreion$ actually leads to more rapidly
rising SFHs, as it allows for more star formation during the 
pre-reionization era, which reduces the relative fraction of stars
that format late times.  Furthermore, because a given galaxy tends 
to form more stars overall with later $\zreion$, some galaxies 
will shift to higher mass bins, further biasing the median SFH of the 
galaxies remaining in the lower mass bin towards earlier times (when
normalizing the SFHs).  We find qualitatively identical trends among 
the centrals around the \run{m12i}, and when performing the same comparison 
with the dwarf galaxies around \run{m12f} (not plotted).

\bsp	
\end{document}